\documentclass[fleqn,usenatbib]{mnras}
\usepackage[T1]{fontenc}
\usepackage{newtxtext,newtxmath}
\DeclareRobustCommand{\VAN}[3]{#2}
\let\VANthebibliography\thebibliography
\def\thebibliography{\DeclareRobustCommand{\VAN}[3]{##3}\VANthebibliography}

\usepackage{booktabs}
\usepackage{ae,aecompl}
\usepackage{xcolor}

\usepackage{graphicx}
\usepackage{amsmath}	

\title[Full-spectrum fitting of SSPs: mass dependence]{On the
  precision of full-spectrum fitting of simple stellar
  populations. II. The dependence on star cluster mass 
  in the wavelength range  0.3\,--\,5.0 $\mu$m} 
\author[Goudfrooij \& Asa'd]{Paul Goudfrooij$^1$\thanks{e-mail: goudfroo@stsci.edu} and Randa
  S.\ Asa'd$^{2,1}$\thanks{e-mail: raasad@aus.edu}  \smallskip \\  
$^1$Space Telescope Science Institute, 3700 San Martin Drive,
  Baltimore, MD 21218, USA \\
$^2$Physics Department, American University of Sharjah, P.O.\ Box
  26666, Sharjah, UAE
}
\date{Accepted 2020 November 2. Received 2020 October 6; in original
  form 2020 August 6}

\begin{document}

\volume{501}
\pagerange{440--467} \pubyear{2021} 

\maketitle

\label{firstpage}

\begin{abstract}
In this second paper of a series on the accuracy and precision of the
determination of age and metallicity of simple stellar populations
(SSPs) by means of the full spectrum fitting technique, we study the
influence of star cluster mass through stochastic fluctuations of the
number of stars near the top of the stellar mass function, which
dominate the flux in certain wavelength regimes depending on the
age. We consider SSP models based on the Padova isochrones, spanning
the age range $7.0 \leq \mbox{log\,(age/yr)} \leq 10.1$. Simulated
spectra of star clusters in the mass range
$10^4 \leq M/M_{\odot} < 10^6$ are compared with SSP model spectra to determine
best-fit ages and metallicities using a full-spectrum fitting routine in four 
wavelength regimes: the blue optical (0.35\,--\,0.70 $\mu$m), the red
optical (0.6\,--\,1.0 $\mu$m), the near-IR (1.0\,--\,2.5 $\mu$m), and
the mid-IR  (2.5\,--\,5.0 $\mu$m). We compare the power of each
wavelength regime in terms of both the overall precision of age and
metallicity determination, and of its dependence on cluster mass. We
also study the relevance of spectral resolution in this context by
utilizing two different spectral libraries (BaSeL and BT-Settl). We
highlight the power of the mid-IR regime in terms of identifying young
massive clusters in dusty star forming regions in distant galaxies.
The spectra of the simulated star clusters and SSPs are made available
online to enable follow-up studies by the community.   
\end{abstract}
\begin{keywords}
stars: abundances -- stars: luminosity function, mass function -- star clusters: general 
\end{keywords}

\section{Introduction}
\label{s:intro}

Star clusters are generally considered to be the best known examples of simple
stellar populations (SSPs). They constitute compact stellar systems containing
thousands to millions of stars with essentially the same age and
metallicity. Star clusters are also highly relevant for galaxy formation and 
  evolution studies, since a significant fraction of star formation occurs in clusters
  \citep[e.g.,][]{Lada03}.
Since clusters can be observed with useful signal-to-noise ratios out
to distances of tens to hundreds of Mpc, they are key to providing important
clues of the star formation and chemical enrichment history of their host
galaxies by means of measurements of their ages, metallicities, and masses
\citep[e.g.,][]{Brodie06,Goudfrooij07,Goudfrooij12,Fall09,Chandar10,
  Chilingarian18}.
Observational studies of cluster populations in galaxies depend
  critically on two basic limitations and how to deal with them: 
  (1) observational biases in terms of photometric and spatial completeness
  fractions of various cluster samples. For example, young cluster populations
  tend to be found in star formation regions with strong and spatially variable
  amounts of extinction, which causes a wavelength dependence to completeness
  fractions which impact the low-luminosity end of cluster luminosity functions; 
  (2) the limited accuracy and precision of methods to estimate cluster ages,
  metallicities, and masses.

With the exception of star clusters in galaxies within the Local Group, one has
to use integrated-light measurements to estimate their properties. This
estimation involves the use of synthetic stellar population synthesis (SPS) models,
which generally assume that the initial mass function (IMF) of stellar systems
is uniformly populated, i.e., that the distribution of stellar masses is
continuous and that all the evolutionary stages are well sampled. However, this
assumption is incorrect for real star clusters due to their finite mass. Hence,
the level of validity of a comparison between the predictions of synthesis
models and observations of real clusters depends on the cluster mass.  
This phenomenon is often referred to as ``stochastic fluctuations'' and is due
to the confluence of the small fractional number of bright stars in a stellar
population and their large contribution to the luminosity. Random sampling of
the stellar mass function can thus cause significant changes in the resulting
spectral energy distribution (SED) of star clusters
\citep[e.g.,][]{Girardi93,Santos97,Lancon00}, leading to incorrect
  determinations of cluster ages and masses.  

The effects of stochastic fluctuations on photometric observations of star
clusters have been studied by several teams
\citep[e.g.,][]{Lancon00, Cervino04, Cervino06, Piskunov09, Popescu09,
  Popescu2012, Fouesneau10, Fouesneau12, Fouesneau14, Krumholz15, Krumholz19}. 
These studies have shown that the predicted colour distributions
depend strongly on the cluster mass (especially in the near-IR regime) and that
they can be significantly non-Gaussian,
even for cluster masses exceeding $10^5\; M_{\odot}$.
 Best-fit ages derived by comparing the colours of low-mass mock clusters with
  standard SPS model predictions have been found to concentrate around certain
  discrete ages corresponding to significant and luminous events in stellar
  evolution such as the sudden onset of supergiants, the upper asymptotic giant 
  branch (AGB), and the red giant branch (RGB). This standard method often 
  underestimates cluster ages significantly by up to 1.5 dex, a precision that
  can be improved significantly by using mass-dependent colours produced using
  ``discrete'' population synthesis
  \citep{Popescu10a,Popescu10b,Fouesneau10,Krumholz19}.

In the current paper, we expand the investigation of the effects of stochastic
fluctuations as a function of wavelength to analyze results obtained from the
full-spectrum fitting method, which has become popular through its virtue of
incorporating all features in the SED of stellar systems. While this
  method is more sensitive to the accuracy of flux calibration than the
  other popular method of index fitting using equivalent widths of spectral
  features \citep[e.g.,][]{Worthey94,Schiavon07}, it uses the continuum shape as well as
  all spectral lines, which can significantly improve the precision of age
  determination \citep{Bica86a, Santos95, Santos06, Fernandes10, Sanchez11,
    Benitez-Llambay12, Asad14, Wilkinson17, Chilingarian18}. 

In \citet[][hereafter paper I]{Asad20}, we investigated the precision of age and
metallicity estimation using the method of full-spectrum fitting of
integrated-light spectra as functions of the S/N of the data and the wavelength
range used in the fitting within the optical wavelength regime. In this second
paper of the series, we examine the dependence of the precision of this fitting
technique on the mass of the star cluster, through its effects of random
sampling of the stellar mass function. We perform this analysis for the age
 range $7.0 \leq \mbox{log\,(age/yr)} \leq 10.1$ in four different wavelength
ranges from the optical through mid-IR regimes (0.3\,--\,5.0 $\mu$m), thus
providing relevant information on the power of those wavelength regimes in
  terms of the precision of the determination of ages, metallicities, and masses
  of star clusters. These results are relevant for both optical and near-IR (NIR)
spectroscopy as well as future space-based spectroscopy such as with the NIR
instruments aboard the James Webb Space Telescope (JWST).   

The simulations and methods are described in Section~\ref{s:data}, followed by a
detailed discussion of the results in Section~\ref{s:results}. We summarize our
findings in Section~\ref{s:summ}.  

\section{Data and Methods}
\label{s:data}

\subsection{Mock Star Clusters and their Integrated-light Spectra}
\label{s:mockGCs}

\subsubsection{Methodology}

For the purposes of this paper, all stars in a star cluster are presumed to have
the same age and chemical abundance. This constitutes a simplification of the
real situation, since  most old globular clusters are now known to host
variations of abundances of several light elements (mainly He, C, N, O, Na, and
Al; see  \citealt{BastianLardo18} and references therein). However, these
variations seem to be absent in clusters younger than $\sim$\,2 Gyr
\citep[see][]{Martocchia18}, and the purpose of this paper is to test the
influence of cluster mass on the precision of the determination of ages and
metallicities of integrated-light spectra. As such, we treat star clusters as
SSPs, we assume solar element abundance ratios, and we ignore the influence of
binary stars.  

Star clusters with masses between $10^4\:M_{\odot}$
  and $10^6\:M_{\odot}$ are simulated as described in detail in 
\citet{Goudfrooij09,Goudfrooij11a}. Briefly, we populate Padova isochrones 
from \citet{Marigo08} with stars randomly drawn from a 
\citet{Kroupa01} initial mass function (IMF) between the minimum (0.15
$M_{\odot}$) and the maximum stellar mass in the isochrone table in
question. The total number of stars $N$ in a cluster simulation is determined as
follows:  
\begin{equation}
  N = M_{\rm cl}/ \overline{M_{\rm *}} 
\label{eq:Nstars}
\end{equation}
where $M_{\rm cl}$ is the cluster mass and $\overline{M_{\rm *}}$ is the mean
stellar mass in the isochrone for the \citet{Kroupa01} IMF
\citep[see also][]{Popescu09}. 
For each star in the simulation, values for log $L$, log $T_{\rm eff}$, and
log $g$ are determined by means of linear interpolation between entries in the
relevant isochrone, using the initial stellar mass $M_{\rm *,i}$ as the independent variable.
The number of independent cluster simulations for a given cluster
mass was chosen to produce a grand total of $5\times10^6\;M_{\odot}$ at each
age. We created these simulated clusters for a grid in log\,(age/yr) between 7.0 and
10.1 with a step size $\Delta$\,log\,(age/yr) = 0.1. As in paper I, we adopt a fixed
cluster metallicity of [Z/H] = $-$0.4 for this study (corresponding to the
metallicity of the Large Magellanic Cloud (LMC)). 

Integrated-light spectra are derived for each simulated
cluster by creating and co-adding spectra of all constituent stars, using 
a script \textsc{intspec} which involves M.\ Fouesneau's  
\href{http://mfouesneau.github.io/docs/pystellibs}{{\tt pystellibs}\footnote{http://mfouesneau.github.io/docs/pystellibs}}
tools in \textsc{Python}, featuring 2-D linear interpolation in
$\log\, T_{\rm eff} -  \log\, g$ space and linear interpolation in metallicity. 
To enable a comparison between spectral libraries in terms of model ingredients
as well as spectral resolution, we use two synthetic libraries: \\ [-3.5ex] 
\begin{enumerate}
    \item BaSeL \citep{Lejeune98} which 
comprises a theoretical Kurucz library that was recalibrated using empirical
photometry. This recalibration is defined  for the metallicity range $-1.0 <
\mbox{[Z/H]} < +0.5$ for both dwarf and giant stars.  The spectral resolution of
the BaSeL library is 20\,\AA\ in the wavelength range 2900\,\AA\,--\,1\,$\mu$m,
50\,\AA\ (1.0\,--\,1.6 $\mu$m), 100\,\AA\ (1.6\,--\,3.2 $\mu$m),
200\,\AA\ (3.2\,--\,6.4 $\mu$m), and 400\,\AA\ (6.4\,--\,10.0 $\mu$m).  
\item BT-Settl \citep{Allard12} which is based on the Phoenix code with updates
  for water vapor, methane, ammonia and CO$_2$, as well as a cloud model that
  uses 2-D radiation hydrodynamic simulations. The spectral resolution of the
  BT-Settl library is 2\,\AA\ ($\lambda \le 1.05\;\mu$m), 4\,\AA\ (1.05\,--\,2.5
  $\mu$m), 16\,\AA\ (2.5\,--\,5.4 $\mu$m), 160\,\AA\ (5.4\,--\,30 $\mu$m), and
  0.8 $\mu$m (30\,--\,80 $\mu$m). \\ [-3.5ex] 
\end{enumerate}
The main reasons why we use synthetic libraries in this paper are
  their spectral coverage across the full wavelength range considered here
  (which is particularly relevant to space-based observations) and their uniform
  coverage in log $T_{\rm eff}$ -- log $g$ -- [Z/H] parameter space.  
In this paper, we only consider wavelengths up to 5 $\mu$m, since the BaSeL
library uses blackbody spectra beyond that.  

\subsubsection{Caveats}

It should be noted that the BaSeL and BT-Settl libraries do not include spectra
of carbon stars, which are important in stellar population synthesis of
intermediate-age populations when the contribution of cool,  thermally pulsing
AGB stars to integrated-light spectra is significant
\citep[e.g.,][]{Lancon02,Maraston05,Conroy09}. For example, the number ratio of
carbon stars to M-type (i.e., oxygen-rich) AGB stars can be used as an indirect
indication of the metallicity at the formation era
\citep[e.g.,][]{Cioni05}. The omission of carbon stars in stellar libraries will
therefore have a negative effect on the \emph{absolute} accuracy and precision
of metallicity determination in intermediate-age populations. However, the
actual relation between the C/O number ratio of AGB stars and the metallicity is
still quite uncertain \citep[e.g.,][]{Boyer19}. Moreover, the focus of this
paper is on the effect of \emph{cluster mass}, and the associated stochastic
fluctuations of the distribution of stars across the luminous post-MS stages of
  stellar evolution in isochrones, on the resulting precision of age and metallicity
determination for star cluster spectra. With this in mind, we use the BaSeL and
BT-Settl libraries as is, thus modeling AGB stars as oxygen-rich M-type stars.

Our SED modeling also does not include the effects of circumstellar dust around
cool AGB stars (with $T_{\rm eff} \la 4000$ K). As shown by \citet{Villaume15},
the emission from such dust has an appreciable contribution to the SED 
for $\lambda \ga 4 \mu$m for SSPs with ages in the range 0.2\,--\,5
Gyr. Its contribution generally increases with $\lambda$ (up to $\sim$\,20 $\mu$m),
and decreases with increasing metallicity. However, the effect stays very
moderate at $\lambda \leq 5 \mu$m at the metallicity of the LMC. We will discuss
the possible impact of circumstellar dust emission to our results in
Section~\ref{s:MIR}. 

 Finally, we also do not include blue horizontal branch (BHB) stars in our
  SED modeling. The high $T_{\rm eff}$ of such HB stars in globular clusters (up to
  $\approx$\,30,000 K) has received extensive attention in the literature, and
  is generally thought to be due to two possible causes that aren't well
  understood, and not modeled in standard isochrones: a metallicity-dependent
  mass loss rate on the RGB \citep[e.g.,][]{Greggio90} and/or a cluster
  mass-dependent fraction of stars with high helium abundance
  \citep[e.g.,][]{Brown16,Goudfrooij18,Milone18}.  In the context of this paper,
  the main relevance of hot BHB stars is that their presence increases the
  equivalent widths of hydrogen absorption lines, and thus makes spectra of old
  clusters look younger than they really are \citep{Lee00,Maraston00}. However,
  with adequate signal-to-noise ratio and spectral resolution, the spectral
  signatures of a BHB can actually be recognized in optical spectra
  \citep[e.g.,][]{Schiavon04,Trager05}. Furthermore, direct evidence
  of such ``apparent  youth'' due to BHB stars among globular clusters
  with metallicities in the range considered here
  (i.e., $\mbox{[Z/H]} > -1$) has so far only been found in 
  NGC\,6388 and NGC\,6441 \citep[e.g.,][]{Maraston03}, two of the most massive
  globular clusters in our Galaxy ($\log(M/M_{\odot}) \sim 6.5$), while
  metal-rich globular clusters with masses $\la 10^6\;M_{\odot}$
  do not show evidence for BHB stars \citep[see][]{Milone18}. It 
  therefore seems unlikely that the omission of this effect has a significant
  impact on our results, given the cluster masses considered in this paper.

\begin{figure*}
  \centerline{
    \includegraphics[width=0.95\textwidth]{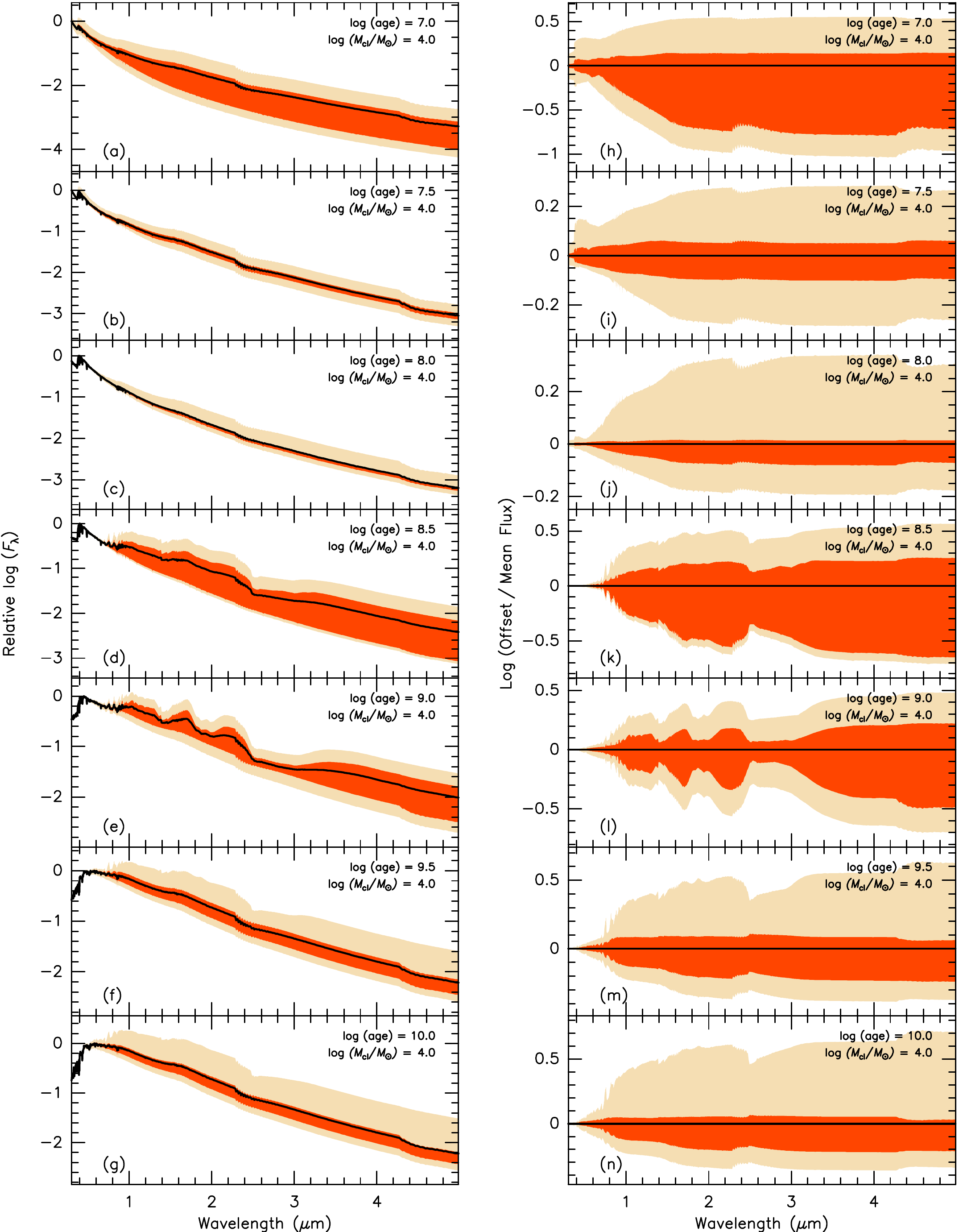}}
  \caption{Star cluster SEDs and the effects of stochastic fluctuations of the
    number of stars near the top of the stellar mass function for a simulated
    cluster mass of $10^4\;M_{\odot}$. The metallicity is [Z/H] = $-0.4$. {\it
      Left panels (a\,--\,g)}: Cluster SEDs in log\,($F_{\lambda}$)
    vs.\ $\lambda$ for log\,(age/yr) = 7.0, 7.5, 8.0, 8.5, 9.0, 9.5, and 10.0
    (see legend), using the BaSeL spectral library. The solid black line
    represents the mean SED of all simulated clusters at that age, normalized by
    its maximum $F_{\lambda}$. The beige shading 
    indicates the full region occupied by the simulated cluster SEDs (using the
    same normalization), while the red shading indicates the region occupied 
    by simulated cluster SEDs between 25\% and 75\% of the full distribution.
    {\it Right panels (h\,--\,n)}: Same as left panels, but now plotted as the
    logarithm of the offset from the mean $F_{\lambda}$ relative to the mean 
    $F_{\lambda}$.
  Note the significant scatter at $\lambda \ga 0.7 \mu\mbox{m}$, which is
  discussed in Section~\ref{s:results}.  
  }
  \label{f:clusspec_1e4}
\end{figure*}
  
\begin{figure*}
  \centerline{
    \includegraphics[width=0.95\textwidth]{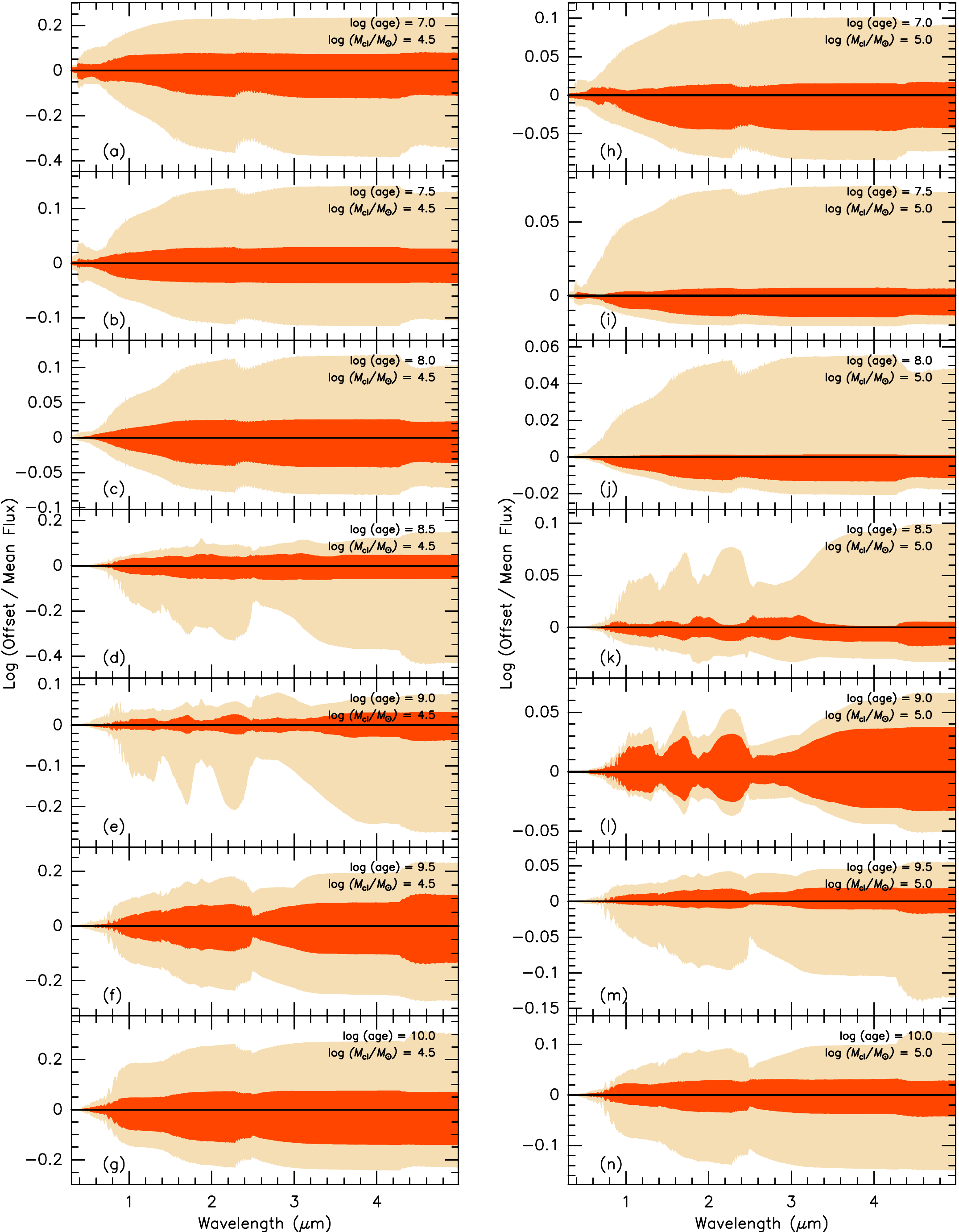}}
  \caption{{\it Left panels (a\,--\,g)}: Same as right panels in
    Figure~\ref{f:clusspec_1e4}, but now for a cluster mass of 
    $3\times10^4\;M_{\odot}$. {\it Right panels (h\,--\,n)}: Same as
    left panels, but now for a cluster mass of 
    $10^5\;M_{\odot}$.}
  \label{f:clusspec_3e4_1e5}
\end{figure*}

\begin{figure*}
  \centerline{
    \includegraphics[width=0.95\textwidth]{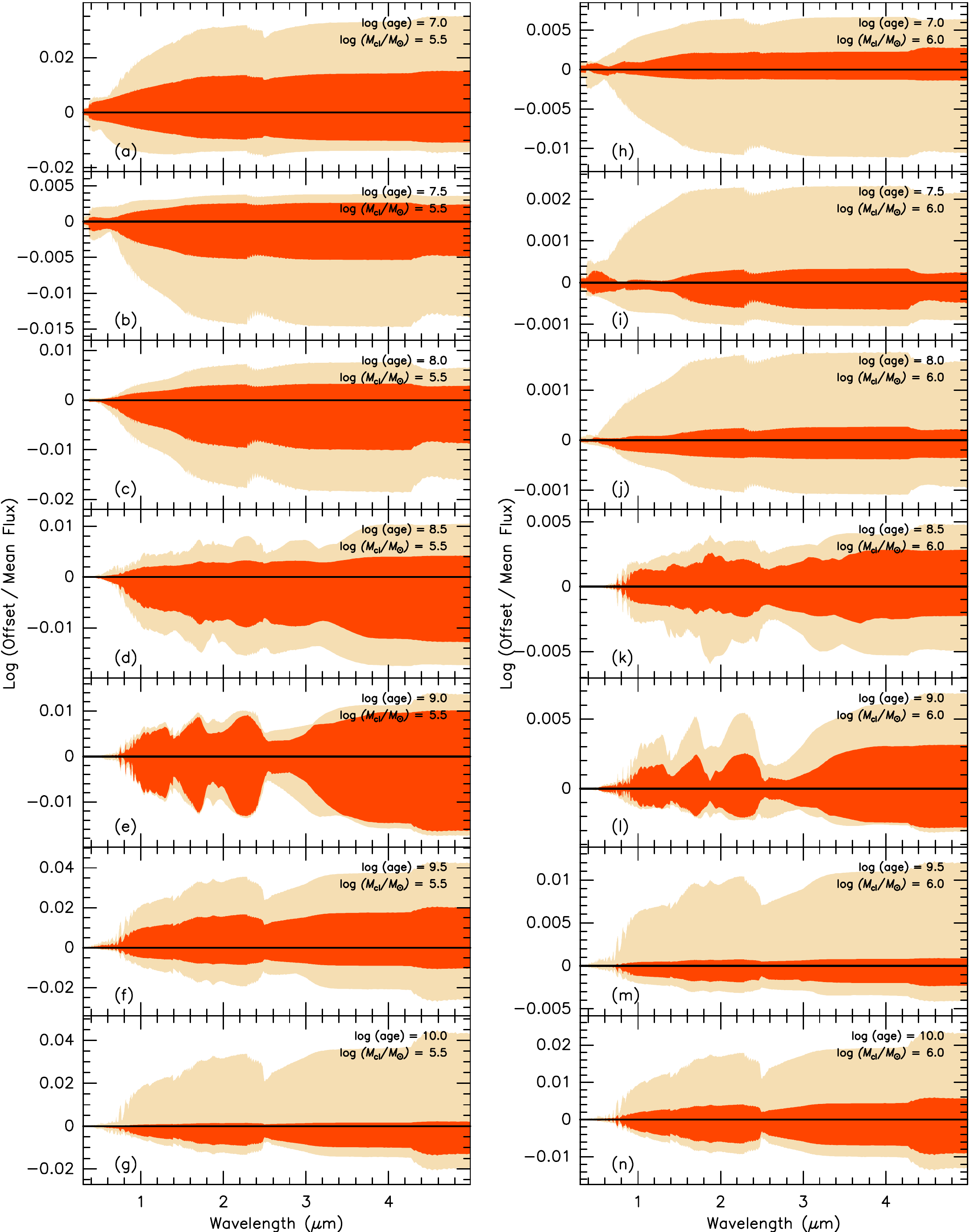}}
  \caption{{\it Left panels (a\,--\,g)}: Same as right panels in
    Figure~\ref{f:clusspec_1e4}, but now for a cluster mass of 
    $3\times10^5\;M_{\odot}$. {\it Right panels (h\,--\,n)}: Same as
    left panels, but now for a cluster mass of 
    $10^6\;M_{\odot}$.}
  \label{f:clusspec_3e5_1e6}
\end{figure*}

\subsection{SSP fitting method}
\label{s:SSPfitting}

To compare the synthetic integrated-light spectra of the simulated clusters
with SSP model predictions, we 
create sets of integrated-light spectra of ``idealized'' SSPs (i.e., clusters of
``pseudo-infinite'' mass) using a script \textsc{sspspec} which is similar to
\textsc{intspec} described above, except that the stars' properties are taken
directly from the isochrone entries. Each spectrum derived for the
isochrone entries is then assigned a weight $W_{i}$ that is proportional to
the number of stars at that initial stellar mass, i.e., to the chosen
IMF. Since isochrone models generally do not sample their entries with an
uniform bin size in initial stellar mass ($dM$), we involve the latter 
in $W_{i}$ as follows:
\begin{equation}
  W_{i} = \frac{N_{i}(M-dM/2, \, M+dM/2)}{N_{\rm total}} =
  \frac{\int_{M-dM/2}^{M+dM/2}\, \psi_i (M) dM}{\int_{m_{\rm min}}^{m_{\rm
        max}} \; \psi_i (M)  M dM} 
\label{eq:IMF}
\end{equation}
where $N_{\rm total}$ is the total number of stars in the SSP, $\psi(M) = dN/dM$
is the IMF, and $m_{\rm min}$ and $m_{\rm max}$ are the minimum 
and maximum initial stellar mass in the isochrone, respectively.  
The integrations are performed numerically using adaptive quadrature. 
For the purpose of this paper, we use the same IMF, isochrone family, and
spectral library as the mock cluster spectra. We create
these SSP model spectra for the age range 6.8 $<$ log\,(age/yr) $<$ 10.2, again
with a grid step size of 0.1 dex. This grid of SSP spectra is created for
metallicities [Z/H] = $-$1.0, $-$0.8, $-$0.6, $-$0.4, $-$0.2, 0.0, and
+0.2, thus staying within the range in which the BaSeL library was
recalibrated using empirical photometry. All model spectra of mock
clusters and SSPs created for this paper are made available in the
online supplementary material of this article for follow-up studies by
the community.

After converting both the mock cluster spectra and the SSP model spectra
to a common flux normalization, we use the full-spectrum fitting
routine {\tt ASAD$_{\tt 2}$} described in \citet{Asad14} and \citet{Asad13,Asad16}
 to obtain the best age and metallicity estimation by minimizing the following quantity: 

\begin{equation} 
\sum_{\lambda=\lambda_{\rm start }}^{\lambda_{\rm end}}
\frac{[(CF)_{\lambda} - (MF)_{\lambda}]^{2}}{(CF)_{\lambda_{\rm norm}}}\,\mbox{,}  
\label{eq:fitchi2}
\end{equation} 
where $CF$ is the mock cluster spectrum, $MF$ is the SSP model
spectrum, and  $\lambda_{\rm norm}$ is the wavelength where $CF$ and
$MF$ are normalized to unity. 

\section{Results and Discussion}
\label{s:results} 

\begin{figure*}
\centerline{\includegraphics[width=0.95\textwidth]{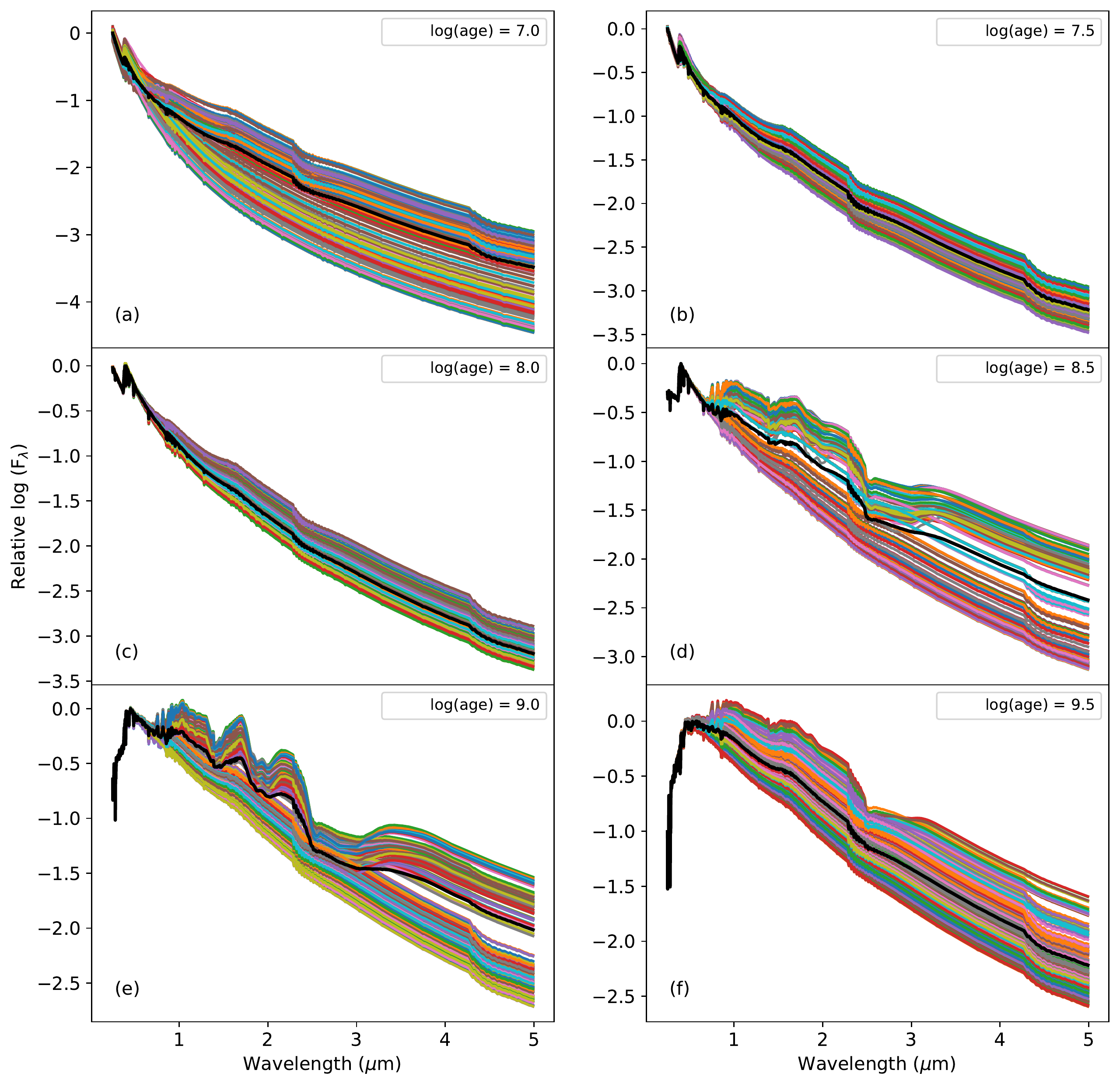}}
\caption{Similar to Figure~\ref{f:clusspec_1e4}, except that we now plot all 500
  individual simulated $10^4\: M_{\odot}$ cluster spectra for the log\,(age/yr)
  values shown in the legends. The thick black line in each panel again
  represents the mean SED of all simulated clusters at that age, normalized by
  its maximum $F_{\lambda}$. All individual cluster spectra were divided by that
  normalization factor. See discussion in Section~\ref{s:results}.}  
\label{f:specplot_all}
\end{figure*}

Figures~\ref{f:clusspec_1e4}\,--\,\ref{f:clusspec_3e5_1e6} illustrate the extent
of variation in the 0.3\,--\,5 $\mu$m SED of SSPs introduced by the stochastic
effects of a limited cluster mass for various ages and cluster masses between
$10^4$ and $10^6\: M_{\odot}$. For the case of $10^4\:M_{\odot}$ clusters, we
show the level of variation both in units of log\,($F_{\lambda}$) and
log\,($F_{\lambda}/\overline{F_{\lambda}}$), where $\overline{F_{\lambda}}$ is
the mean $F_{\lambda}$ of all simulated clusters of a given cluster mass. (For
the higher cluster masses, only log\,($F_{\lambda}/\overline{F_{\lambda}}$ is
shown.) A glance at these Figures immediately highlights a number of general
properties of the effects of stochastic fluctuations as a function of
wavelength: (1) the SED variations generally start to increase significantly
beyond $\lambda \sim 7000$ \AA; (2) the distribution of the variations generally
is asymmetric and non-Gaussian, and the level of its asymmetry varies
significantly with SSP age; and (3) the amplitude of the variations decreases
strongly at strong molecular spectral features at ages in the approximate range
$8.5 \la \mbox{log\,(age/yr)} \la 9.0$.  
To help understand these properties, we plot all 500 individual simulated
spectra for clusters with $M/M_{\odot} = 10^4$ for several ages in
Figure~\ref{f:specplot_all}.  

\begin{figure*}
\centerline{\includegraphics[width=0.93\textwidth]{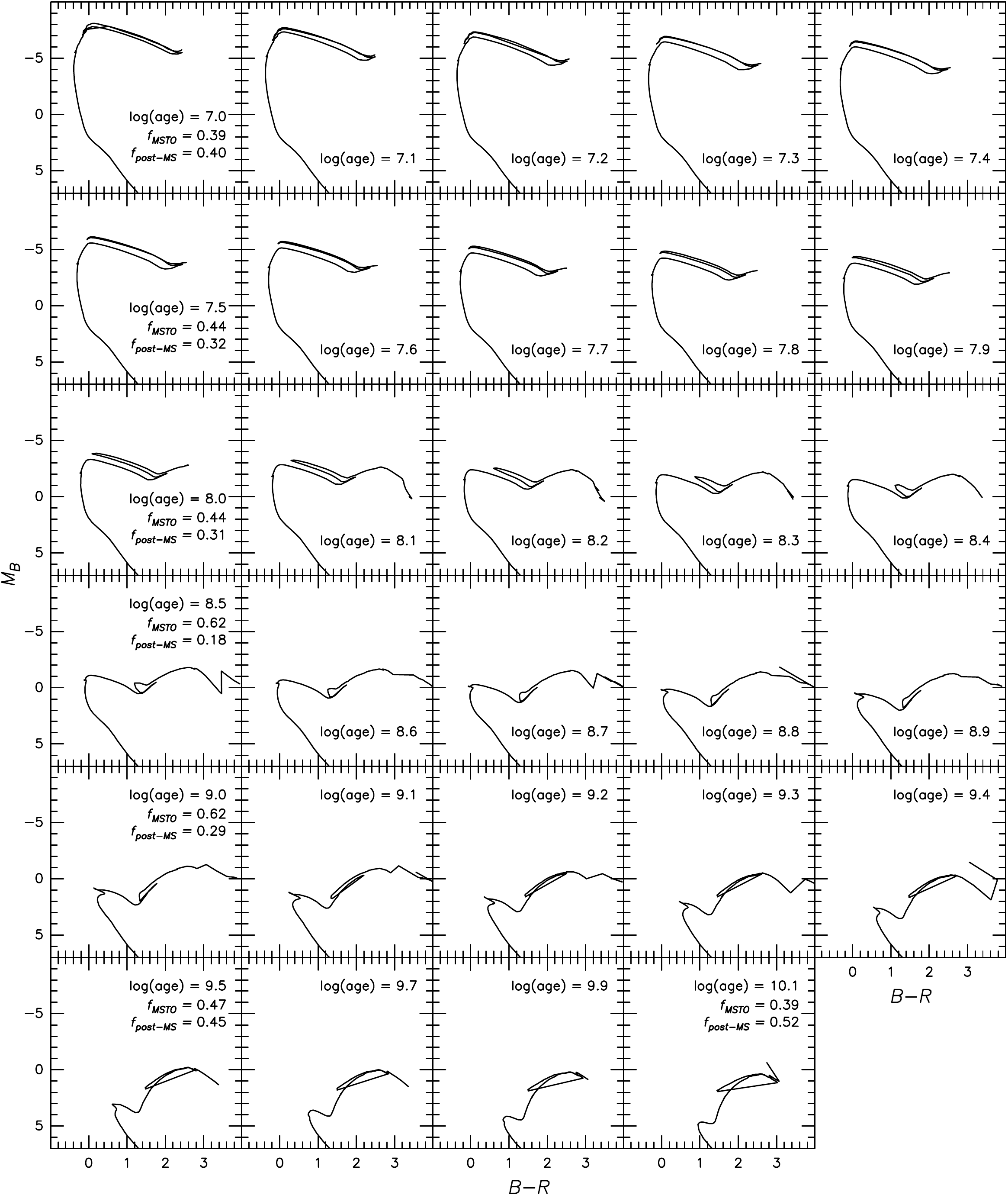}}
\caption{Blue optical range: $B$ vs.\ $B-R$ CMDs for Padova isochrones
  with [Z/H] = $-0.4$ and $7.0 \leq \mbox{log\,(age/yr)} \leq
  10.1$ (see legend in each panel). For log\,(age/yr) = 7.0, 7.5,
8.0, 8.5, 9.0, 9.5, and 10.1, the legend also shows the values of
$f_{\rm MSTO}$ and $f_{\rm post-MS}$, which are the $B$-band luminosity
fractions of MSTO stars and the post-MS stars, respectively, for a
Kroupa IMF. See Section~\ref{s:optical} for details.}
\label{f:cmds_opt}
\end{figure*}

\begin{figure*}
\centerline{\includegraphics[width=0.93\textwidth]{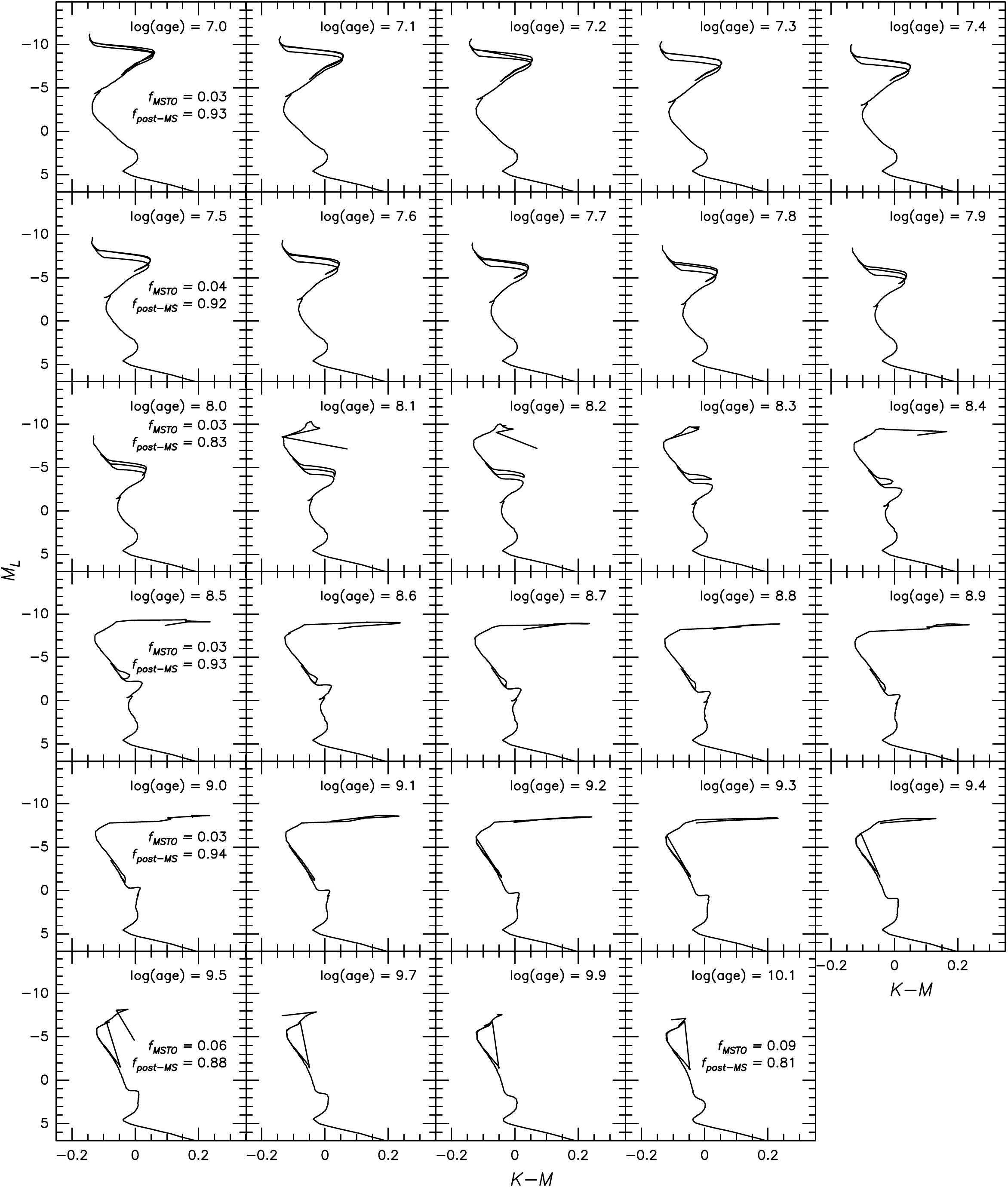}}
\caption{Similar to Figure~\ref{f:cmds_opt} but now for the
    mid-IR range: $L$ vs.\ $K-M$ CMDs for Padova isochrones with
  [Z/H] = $-0.4$ and $7.0 \leq \mbox{log\,(age/yr)} \leq 10.1$ (see
  legend in each panel). Values for $f_{\rm MSTO}$ and $f_{\rm post-MS}$ now
  refer to $L$-band luminosities.}  
\label{f:cmds_midIR}
\end{figure*}

For the remainder of this section, we split the discussion into four wavelength
regimes: the blue optical (3500\,--\,7000 \AA), the red optical (0.6\,--\,1.0
$\mu$m), the NIR (1.0\,--\,2.5 $\mu$m), and the mid-IR (2.5\,--\,5.0
$\mu$m). This choice is based on the results seen in
Figures~\ref{f:clusspec_1e4}\,--\,\ref{f:specplot_all} as well as to highlight
the differences in the power of full-spectrum fitting in terms of age and
metallicity determination in those different wavelength regimes.  
 To set the context for this discussion, we plot colour-magnitude diagrams
  (CMDs) involving filter passbands in these four wavelength regimes in
  Figures~\ref{f:cmds_opt}\,--\,\ref{f:cmds_midIR}. Each of these Figures shows
  a grid of CMDs covering the full age range considered in this paper, thus
  depicting the relative luminosities of the various stages of stellar
  evolution as a function of age. 
  For a subset of ages, Figures~\ref{f:cmds_opt}\,--\,\ref{f:cmds_midIR} also
  list the luminosity fractions produced by main sequence turnoff (MSTO)
  stars and post-MS stars for a Kroupa IMF, denoted by $f_{\rm MSTO}$ and
  $f_{\rm post-MS}$, respectively. In the calculation of $f_{\rm MSTO}$, we
  include stars from the tip of the MSTO down to 1 mag below the turnoff point
  on the CMD in this context. These luminosity fractions are relevant in the
  discussion of the impact of stochastic fluctuations for different ages and
  wavelength regimes.   

\begin{figure*}
\centerline{\includegraphics[width=0.93\textwidth]{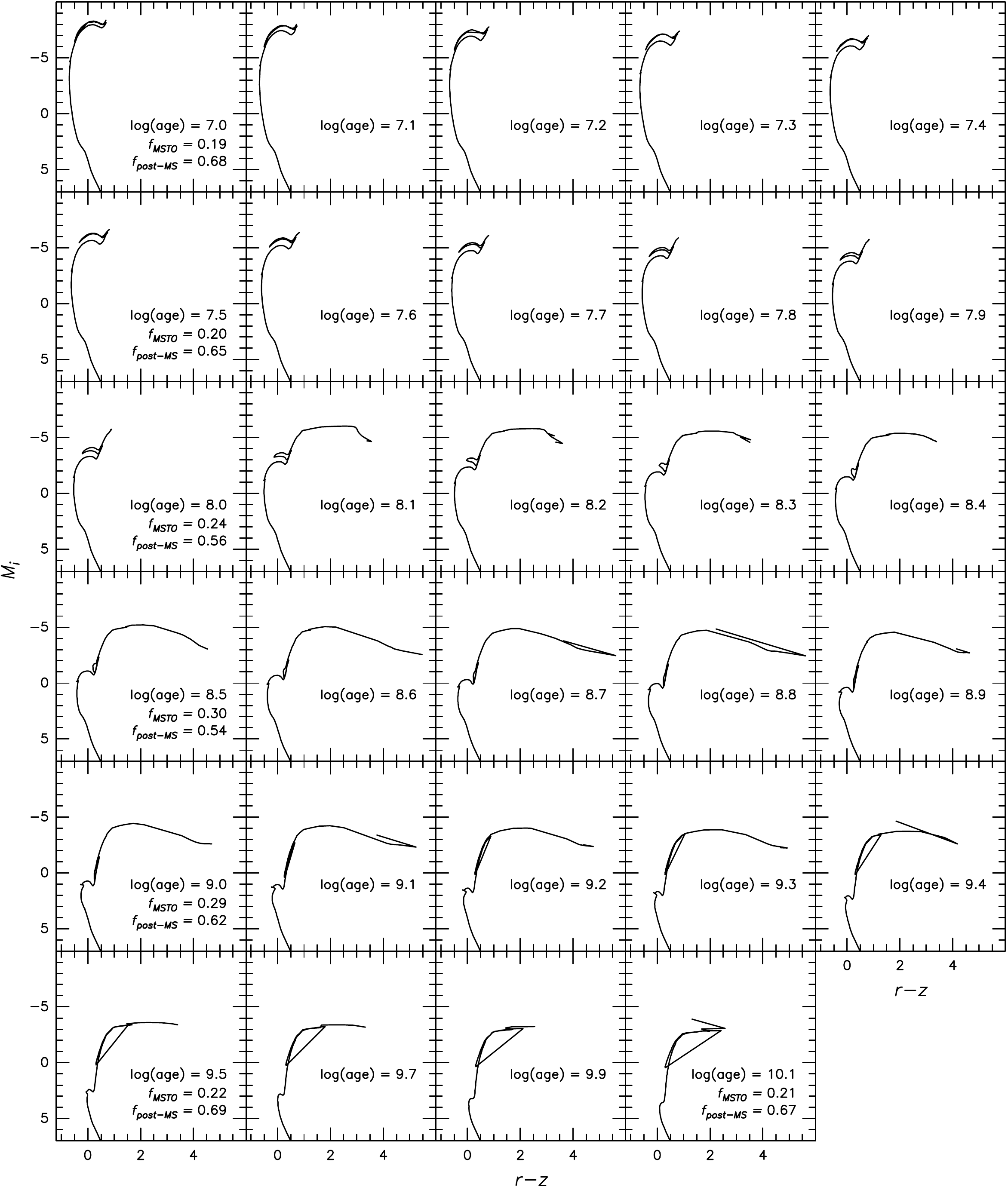}}
\caption{Similar to Figure~\ref{f:cmds_opt} but now for the red
    optical range: $i_{\rm SDSS}$ vs.\ $(r-z)_{\rm SDSS}$ CMDs for
    Padova isochrones with [Z/H] = $-0.4$ and $7.0 \leq
    \mbox{log\,(age/yr)} \leq 10.1$ (see legend in each panel). Values for
    $f_{\rm MSTO}$ and $f_{\rm post-MS}$ now refer to $i_{\rm SDSS}$-band luminosities}. 
\label{f:cmds_red} 
\end{figure*}

\begin{figure*}
\centerline{\includegraphics[width=0.93\textwidth]{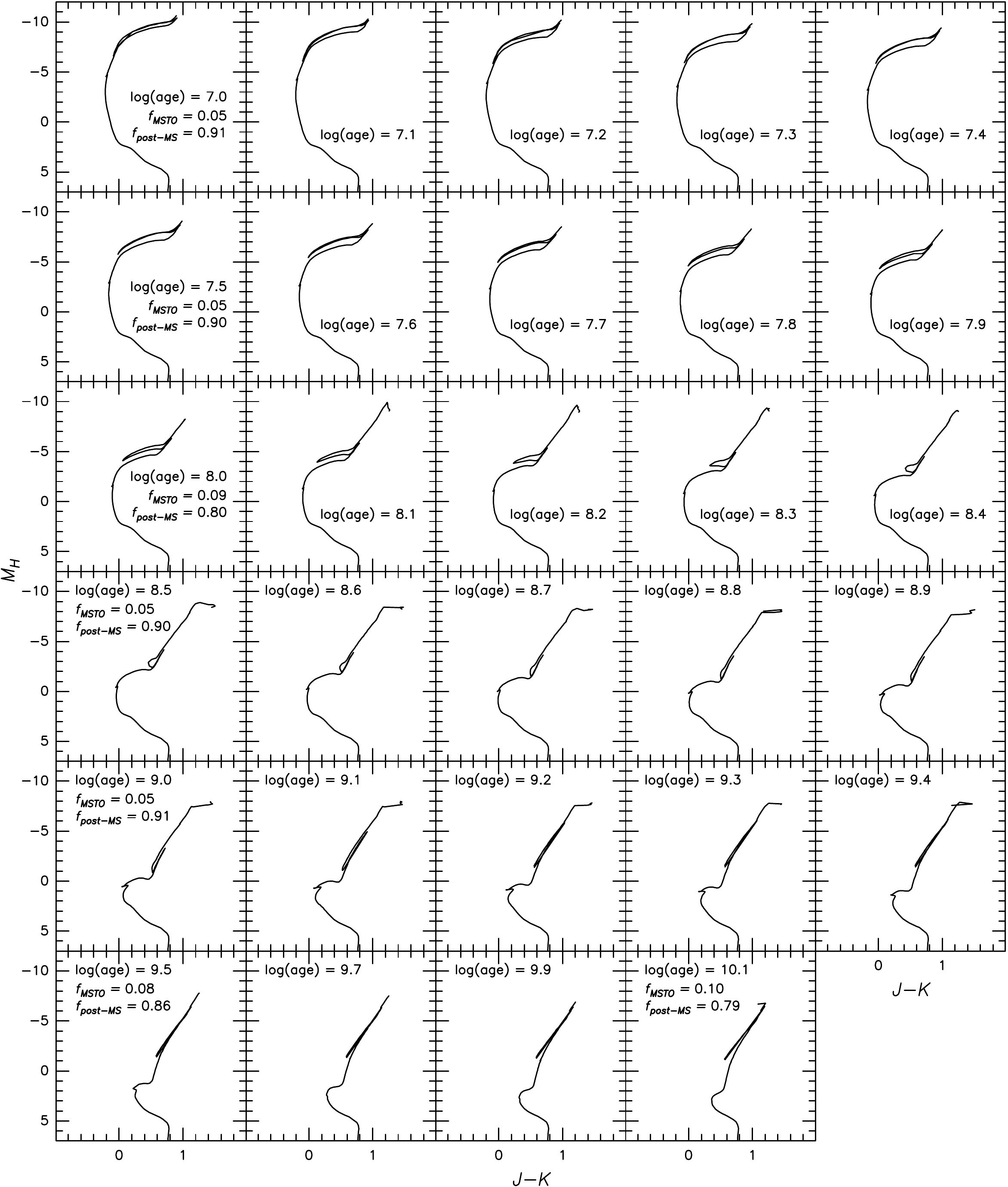}}
\caption{Similar to Figure~\ref{f:cmds_opt} but now for the NIR
    range: $H$ vs.\ $J-K$ CMDs for Padova isochrones with [Z/H] =
  $-0.4$ and $7.0 \leq \mbox{log\,(age/yr)} \leq 10.1$ (see legend in
  each panel). Values for $f_{\rm MSTO}$ and $f_{\rm post-MS}$ now refer to
  $H$-band luminosities.}  
\label{f:cmds_NIR}
\end{figure*}

\begin{figure*}
\centerline{
\includegraphics[width=0.48\textwidth]{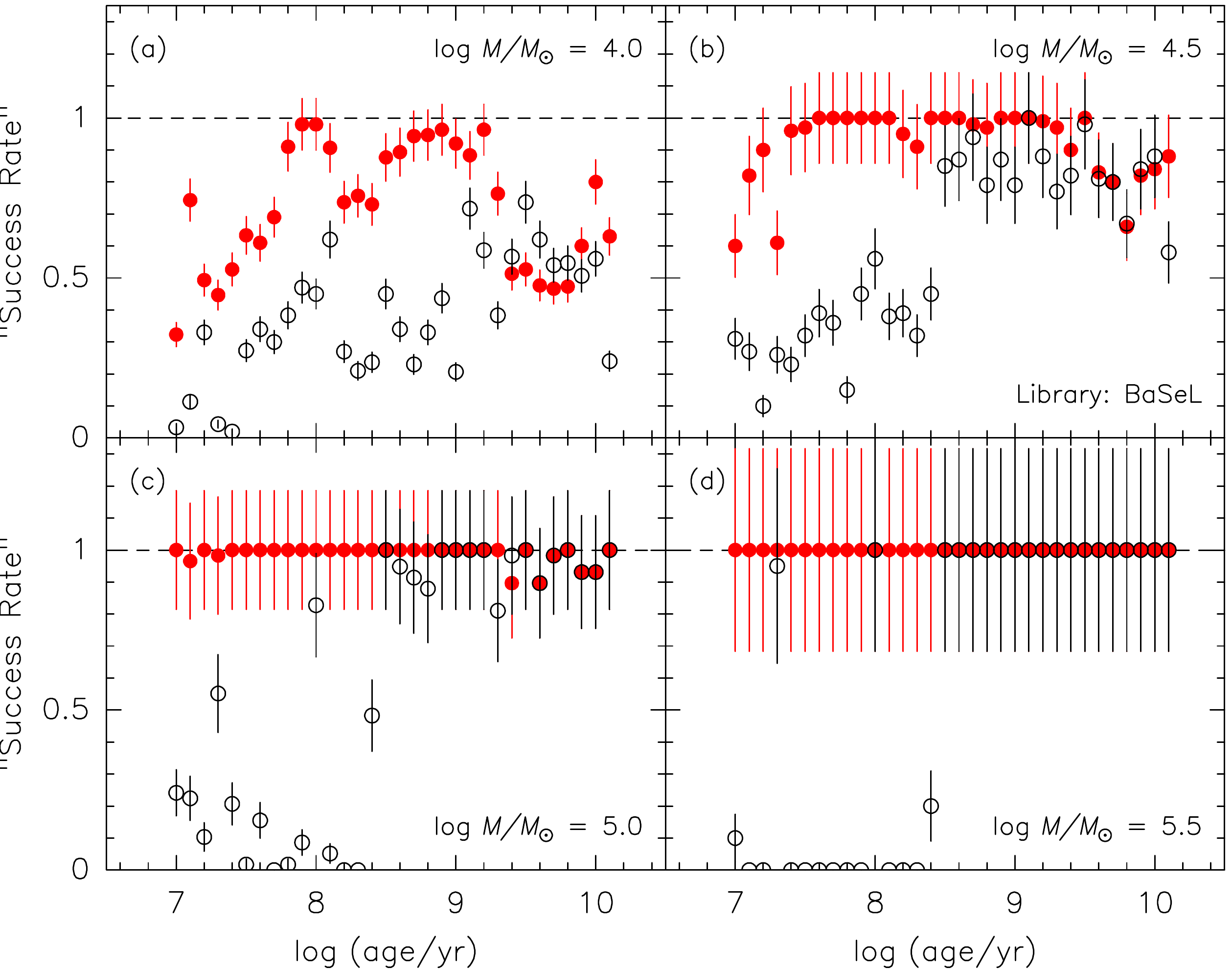}
\hspace*{1mm}
\includegraphics[width=0.48\textwidth]{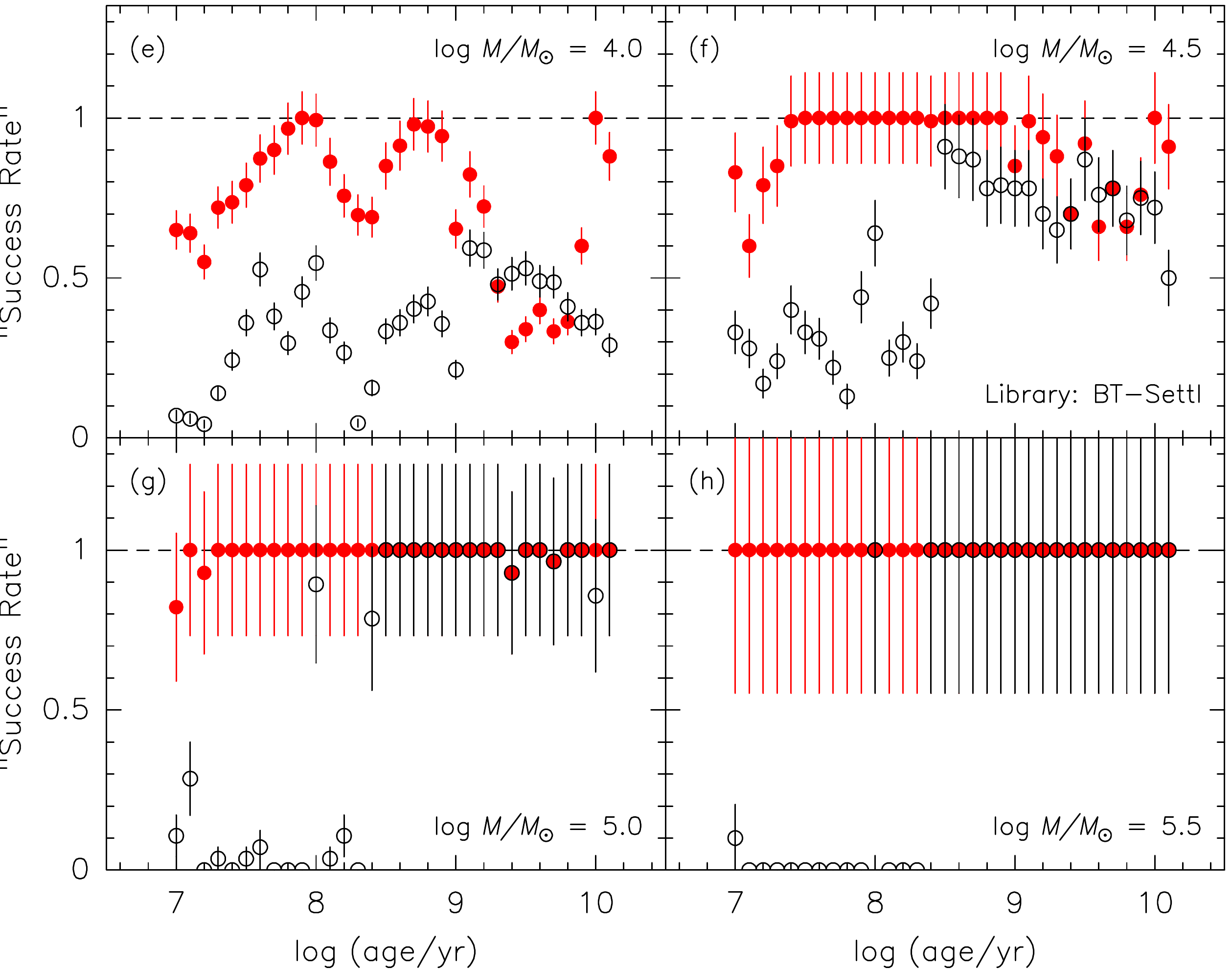}
}
\caption{Blue optical range: Percentage of correctly determined log\,(age)
  (solid red circles) and [Z/H] (open black circles) to within 0.1 dex versus
  log\,(age). Error bars represent Poisson uncertainties. Panels (a)\,--\,(d)
  depict results for the BaSeL library for the cluster masses shown in the
  legend of each panel. Panels (e)\,--\,(h) show the same as panels
  (a)\,--\,(d), but now for the BT-Settl library.} 
\label{f:fractions_age_ZH_opt}
\end{figure*}

\begin{figure}
\centerline{\includegraphics[width=8.cm]{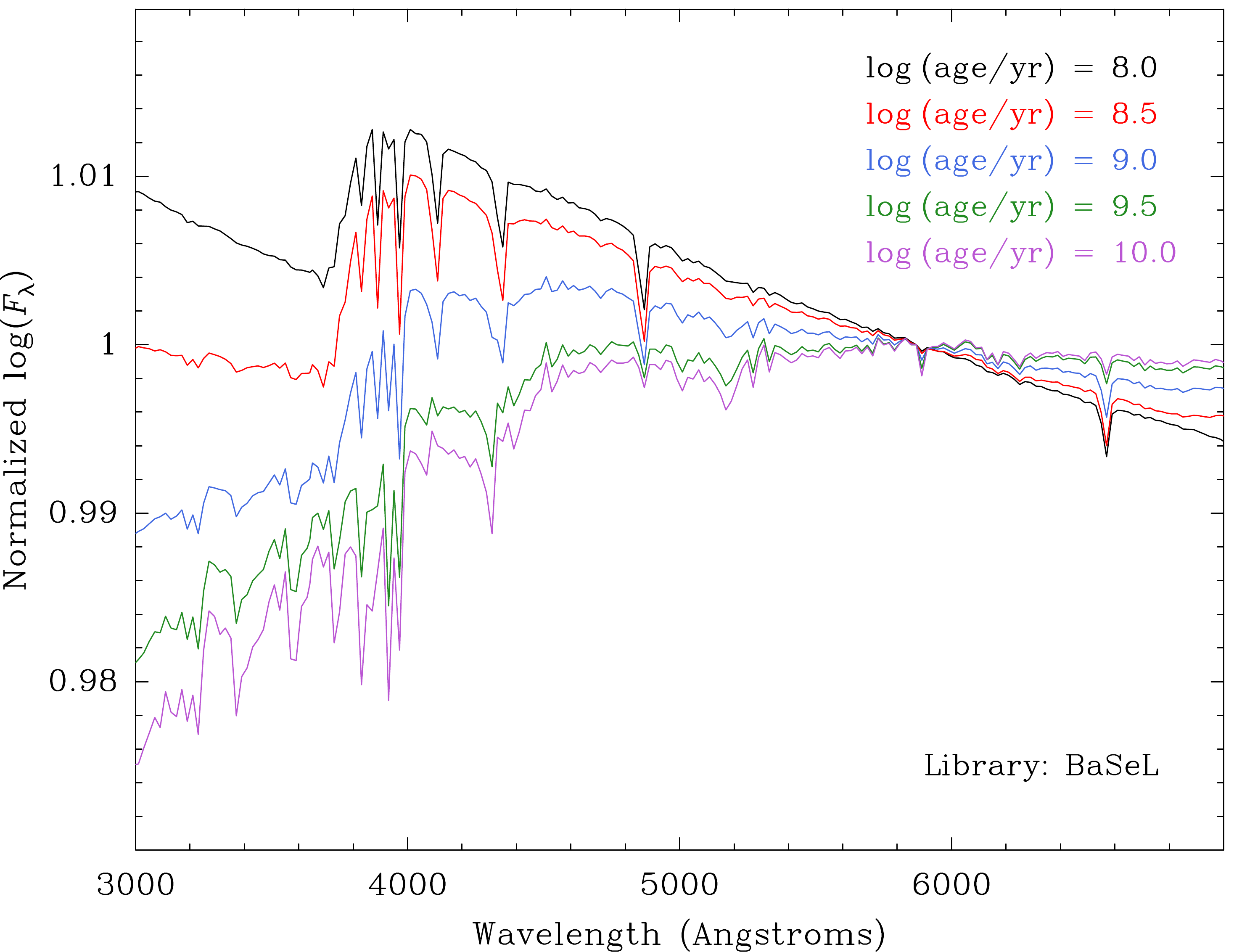}}
\caption{Blue optical SEDs of synthetic SSPs with [Z/H] = $-$0.4. Ages and
  associated line colours are shown in the legend. The SEDs are normalized to
  unity at 5870 \AA.} 
\label{f:opt_SEDs}
\end{figure}

\subsection{Blue Optical Range: 3500\,--\,7000 \AA}
\label{s:optical}

The shape of cluster SEDs in this wavelength range is relatively
insensitive to the stochastic fluctuations of the sampling of
high-mass stars in the stellar MF, when compared with the situation at
longer wavelengths. The main reason for this can be seen in $B$
vs.\ $B-R$ CMDs of Padova isochrones, shown in
Figure~\ref{f:cmds_opt}.
Note that the MSTO region, which is much more densely populated with stars
than the post-MS stages of stellar evolution,
accounts for a high luminosity fraction in this wavelength
range for almost all ages relative to the situation at longer
  wavelengths (cf.\ Figures~\ref{f:cmds_red}, \ref{f:cmds_NIR}, and
  \ref{f:cmds_midIR}).   
And indeed, as shown in the right-hand panels of
Figure~\ref{f:clusspec_1e4} for the simulated clusters with the lowest
considered mass of $10^4\: M_{\odot}$, we find that even for the ages
that show the strongest variation of SED shape (i.e., log\,(age)
$\leq$ 7.5), the FWHM values of the distributions of differences
relative to the true SSP SED stay within 10\%, gradually decreasing
with increasing cluster mass and dropping below 1\%\ at
$M \sim 10^5\: M_{\odot}$.  

We now address how this translates into the precision of age and
metallicity determination using full-spectrum fitting in the blue
optical range as a function of cluster mass. We normalize the spectra
in this wavelength range following Paper 1, using $\lambda_{\rm norm}$
= 5870 \AA\ (cf.\ Equation~\ref{eq:fitchi2}) during the fitting
procedure.  For the purpose of quantifying fitting results, we define
the ``success rate'' as the fraction of clusters whose age or
metallicity are recovered correctly to within 0.1 dex.
Figure~\ref{f:fractions_age_ZH_opt} depicts this ``success rate'' as
functions of age and cluster mass, while
Figures~\ref{f:offsets_age_ZH_opt} and \ref{f:offsets_age_ZH_opt_BT}
(in the Appendix) shows the actual distributions of relative fractions
of offsets from true log\,(age) and [Z/H] seen among all the simulated
clusters. For the lowest-mass clusters considered here (i.e., $10^4\:
M_{\odot}$), the success rate ranges between roughly 40\% and 100\%
for log\,(age), depending on the age. Metallicity determination does
not fare as well, ranging between success rates of just a few \% at
the youngest ages and $\sim$\,75\% at ages of $\ga$ 1.5 Gyr.
 This generally poorer recovery of metallicity relative to that of age is
  partly intrinsic to the method of full-spectrum fitting, since the wavelength
  coverage of spectra by stellar continuum is generally larger 
  than that by strong spectral lines, and continuum shape is 
  more sensitive to age than to metallicity \citep[see,
    e.g.,][]{Bica86a,Benitez-Llambay12}.  
The precision of both age and metallicity determination generally improves 
significantly with increasing cluster mass, with the exception of
metallicity at ages $\la$ 300 Myr. The latter is likely due to the
absence of strong metallic lines in the blue optical regime for hot
stars  (see Figure~\ref{f:opt_SEDs}). Finally, a comparison of the
success rates derived for the BaSeL and BT-Settl libraries (see panels
(a)\,--\,(d) vs.\ panels (e)\,--\,(h) of
Figure~\ref{f:fractions_age_ZH_opt}) reveals no significant
differences for the blue optical regime.  

\subsection{Red Optical Range: 0.6\,--\,1.0\,$\mu$m}
\label{s:red}

Relative to the blue optical region, the sensitivity of the shape of the
integrated-light SED of star clusters to the sampling of high-mass stars in the
stellar MF is stronger in the red optical region (0.6\,--\,1.0 $\mu$m).  This is
evident in the right-hand panels of  Figure~\ref{f:clusspec_1e4} and
Figures~\ref{f:clusspec_3e4_1e5}-\ref{f:clusspec_3e5_1e6}, and can be further
explained by the shapes of $i_{\rm SDSS}$ vs.\ $(r-z)_{\rm SDSS}$ CMDs of the
Padova isochrones (see  Figure~\ref{f:cmds_red}). Comparing the latter with its
counterpart for the blue optical region (i.e., Figure~\ref{f:cmds_opt}), it can
be seen that the luminosity in the red optical region is significantly more
dominated by post-MS stages of stellar evolution, especially the helium-burning
stages (e.g., red supergiants or AGB stars) for which the effects of random
sampling of the stellar MF are generally strongest.  
Figure~\ref{f:specplot_all} illustrates this in that the deviation among 
  SEDs for clusters of $10^4\:M_{\odot}$ strongly rises beyond $\lambda \sim 0.7
  \mu$m. 

\begin{figure*}
\centerline{
\includegraphics[width=0.48\textwidth]{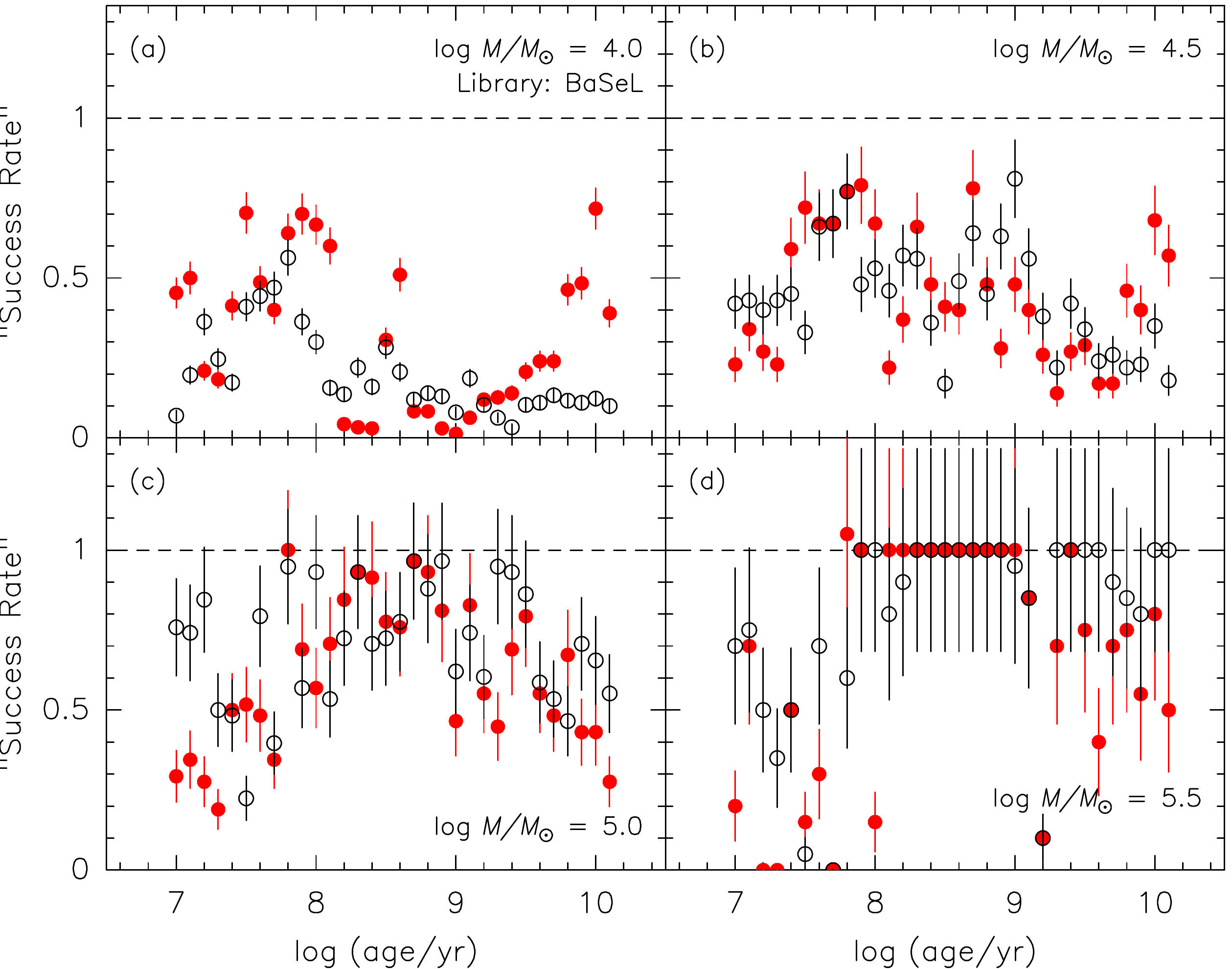}
\hspace*{1mm}
\includegraphics[width=0.48\textwidth]{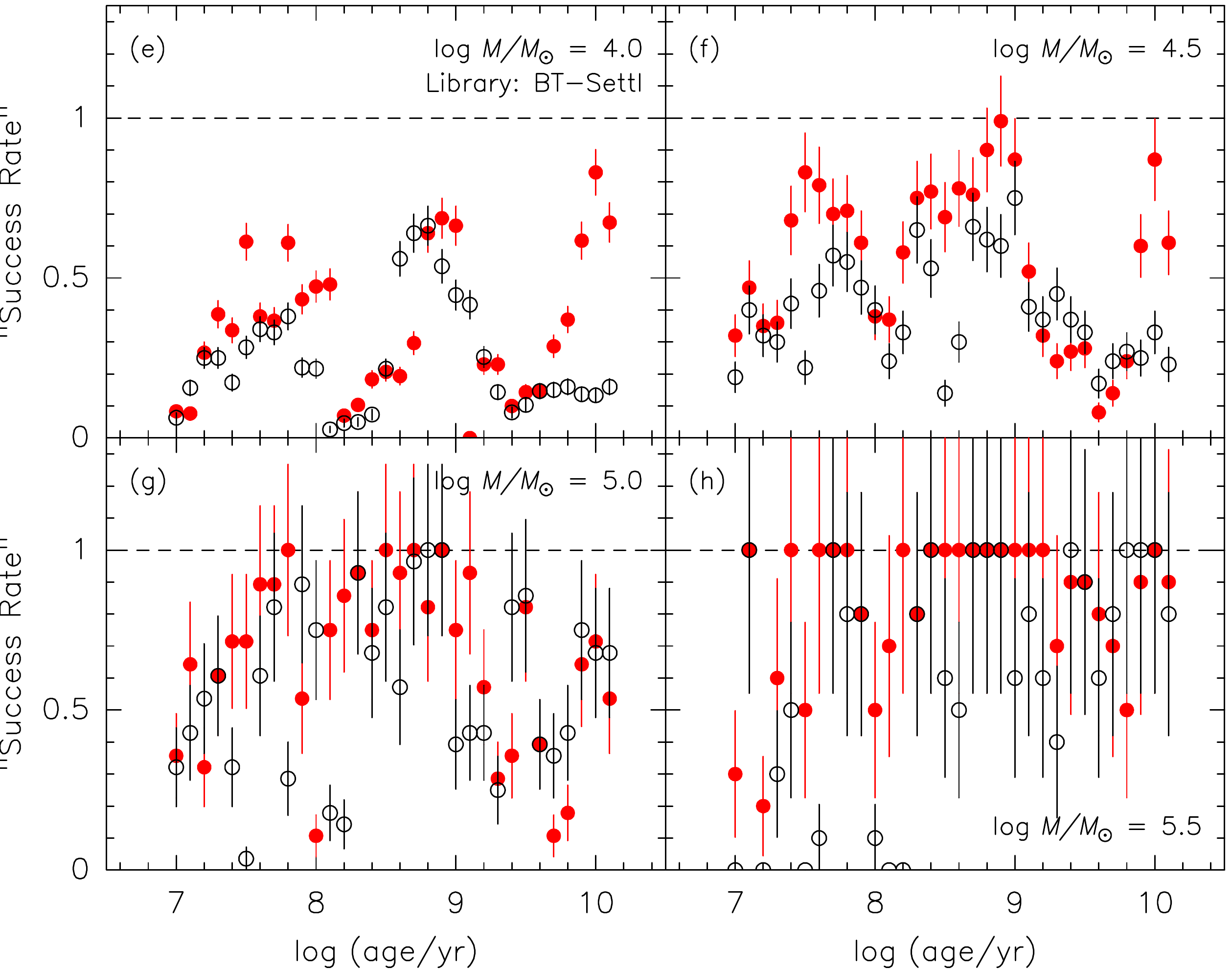}
}
\caption{Same as Figure~\ref{f:fractions_age_ZH_opt}, but now for the red optical range.}
\label{f:fractions_age_ZH_red}
\end{figure*}

The precision of age and metallicity determination in this wavelength regime as
a function of cluster mass is depicted in Figures~\ref{f:fractions_age_ZH_red},
\ref{f:offsets_age_ZH_red}, and \ref{f:offsets_age_ZH_red_BT} (the latter two in
the Appendix). We use  $\lambda_{\rm norm}$ = 9990 \AA\ for the full-spectrum
fitting in this context.  

\begin{figure}
\centerline{\includegraphics[width=8.cm]{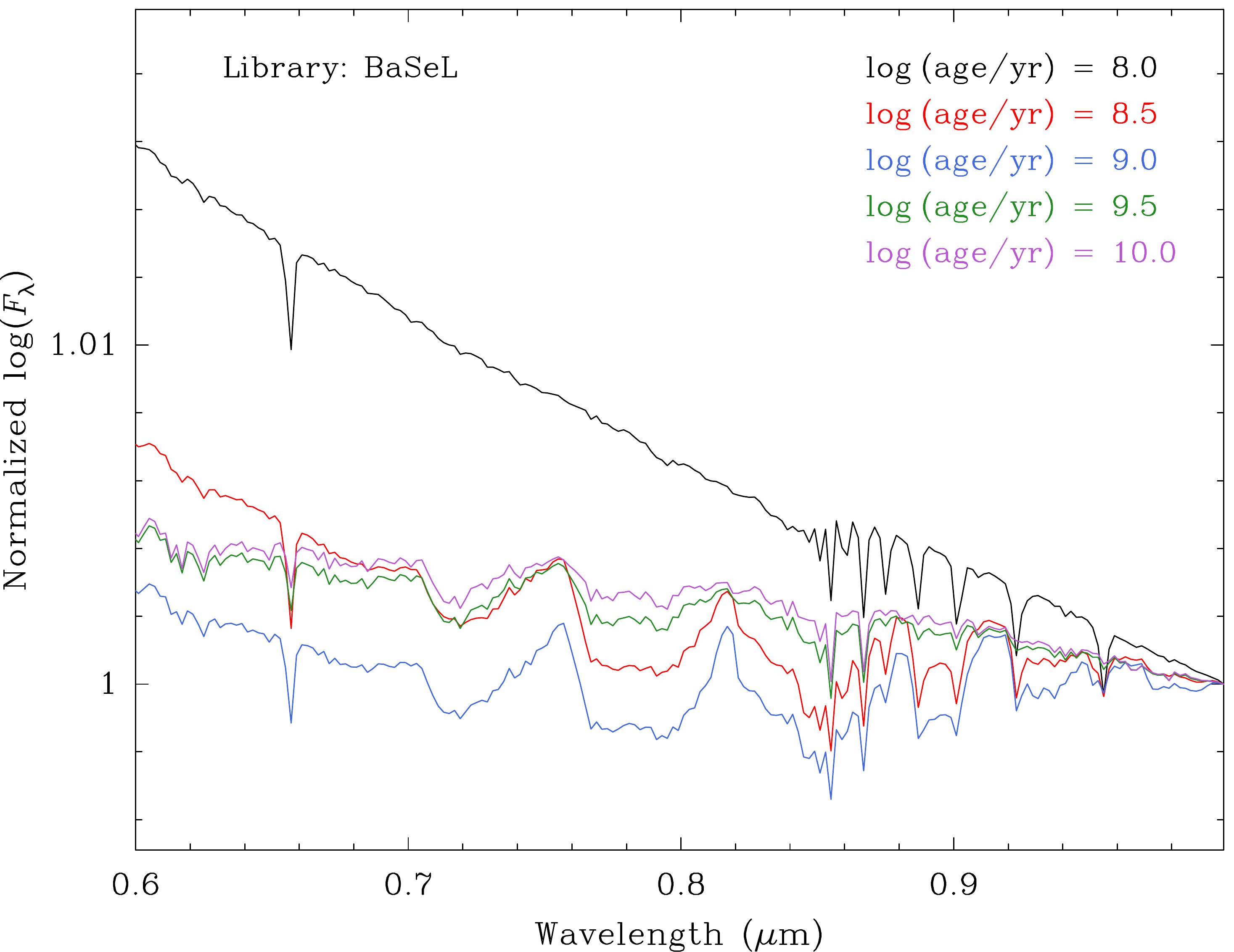}}
\caption{0.6\,--\,1.0 $\mu$m SEDs of synthetic SSPs with [Z/H] = $-$0.4. Ages
  and associated line colours are shown in the legend. The SEDs are normalized to
  unity at 999 nm. Note the strong molecular bands of TiO at $\sim$\,0.72, 0.78,
  0.85, 0.89, and 0.94 $\mu$m for the SSPs with log\,(age/yr) = 8.5 and 9.0,
  when the contribution of TP-AGB stars is largest (see also
  \citealt{LanconWood00}).} 
\label{f:red_SEDs}
\end{figure}

For a cluster mass of $10^4\: M_{\odot}$, we find that the success rate ranges
between just a few \% and 70\% for log\,(age), depending strongly on the
age. The success rate for age determination increases with increasing cluster
mass, as expected, but the rate of this increase is significantly lower than
that seen for the blue optical region in the previous section.  The success rate
of age determination only reaches close to 100\% at a cluster mass of $\sim
3\times10^5\: M_{\odot}$, and only in the age range of $8.1 \la
\mbox{log\,(age/yr)} \la 9.1$. This can be understood by the shape of SSP SEDs
in the red optical region (see Figure~\ref{f:red_SEDs}). For log\,(age) $\la$
8.0, the features in the SEDs are dominated by hydrogen lines (H$\alpha$ and the
higher-order Paschen lines), which are produced by hot stars. However, for
low-mass clusters, the stochastic fluctuations often cause the hot stars in
young clusters not to be dominant in the red optical region (see panel (a) in
Figure~\ref{f:specplot_all}), thus lowering the precision of age determination
significantly.  
For SSPs with $8.1 \la \mbox{log\,(age/yr)} \la 9.1$, the SEDs in the red
optical region are dominated by strong molecular bands of TiO due to the large
contribution from cool TP-AGB stars, which increases with age in this age
interval, thus improving the precision of age determination. The latter then
decreases somewhat towards older ages, when the strength of the TiO bands
decreases significantly and the main sensitivity to age becomes restricted to
the strength of hydrogen lines, which again are produced by the hottest stars in
the clusters, whose contribution to the flux in the red optical region depends
significantly on stochastic fluctuations at these ages (see panel (f) in
Figure~\ref{f:specplot_all}).  

The success rate of metallicity determination in the red optical range behaves
similarly to that of the age determination, both in terms of its overall level
and its dependence on cluster mass, except that it typically does not reach
beyond $\sim$\,80\% for ages $<$ 100 Myr. This is due in part to the lack of
strong metallic spectral features in hot star spectra, especially in this
wavelength region (see Figure~\ref{f:red_SEDs}).  
 
Finally, the age and metallicity determination in this regime is on
average somewhat better with the BT-Settl library than with the BaSeL
one (see Figure~\ref{f:fractions_age_ZH_red}), especially in case of
[Z/H] determination for the lower cluster masses, where the higher
spectral resolution helps the identification of the relevant metallic
features such as TiO and the Ca\,II triplet at 8498, 8542, and 8662 \AA.  

\subsection{Near-IR Range: 1.0\,--\,2.5 $\mu$m}
\label{s:NIR} 

As illustrated by Figures~\ref{f:clusspec_1e4}-\ref{f:clusspec_3e5_1e6}, the
cluster mass has a very strong effect on the level of variations in the shape of
the integrated-light SED of the cluster at $\lambda \ga 1\mu$m, especially at
log\,(age/yr) $\la$ 7.5 and log\,(age/yr) $\ga$ 8.3. As can be seen in NIR ($H$
vs.\ $J-K$) CMDs of Padova isochrones (see Figure~\ref{f:cmds_NIR}), this is
because stars at the top of the stellar MF dominate the NIR luminosity to a
greater extent than at shorter wavelengths. The impact of this effect on the
precision of age and metallicity determination by full-spectrum fitting in the
NIR region are determined by the interplay of the shape and extent of the branch
or branches of stars dominating the NIR luminosity, and the relative density of
stars within those branches as function of luminosity.  
The result of this interplay is different in different age ranges, which we
discuss and illustrate below in some detail. We use  $\lambda_{\rm norm}$ = 2.2
$\mu$m for the full-spectrum fitting in the NIR region, given the lack of
significant spectral features on the short-$\lambda$ side of the first overtone
band of CO at 2.3 $\mu$m. 

In the youngest clusters (age $\la$ 30 Myr), the shape of the SED strongly
depends on the distribution of stars along the blue loop of helium-burning
stars, which encompasses a relatively large range of $T_{\rm eff}$. As can be
seen in panel (a) in Figure~\ref{f:specplot_all}, the variety of resulting SEDs
causes a wide range of derived best-fit SSP ages and hence a very low success
rate for age determination of low-mass clusters in this age range (see
Figure~\ref{f:fractions_age_ZH_NIR}, and Figures~\ref{f:offsets_age_ZH_NIR} and
\ref{f:offsets_age_ZH_NIR_BT} in the Appendix).  

\begin{figure}
\centerline{\includegraphics[width=6.5cm]{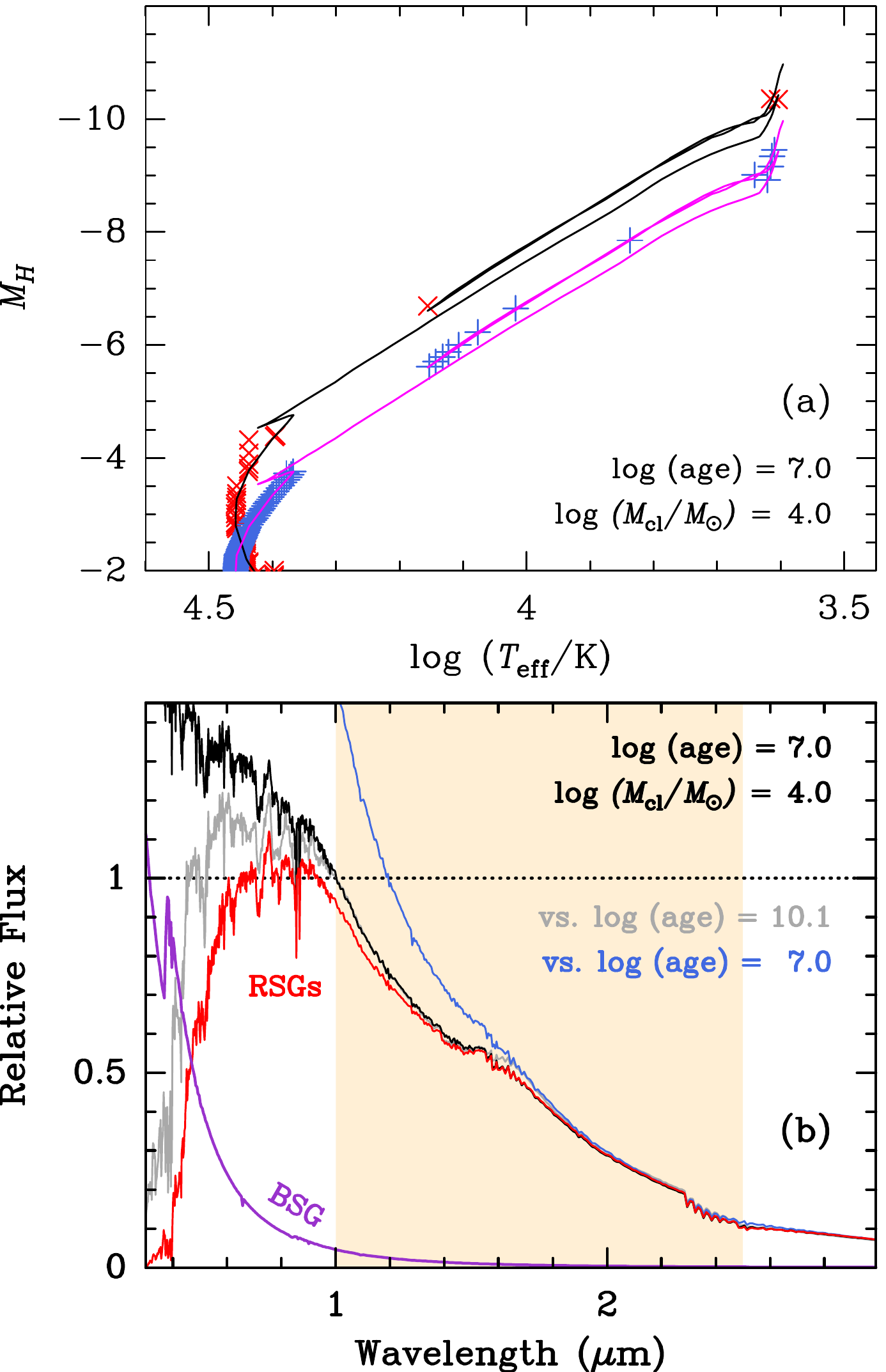}}
\caption{Illustration of simulated $10^4\: M_{\odot}$ cluster with log\,(age/yr)
  = 7.0 masquerading as an SSP with log\,(age/yr) = 10.1 in the NIR. \emph{Panel
    (a)}: $M_H$ vs.\ log\,$T_{\rm eff}$ diagram. The solid black line depicts
  the Padova isochrone with (log\,(age/yr), [Z/H]) = (7.0, $-$0.4). Red crosses
  represent stars in simulated cluster \#111. For comparison, the magenta line
  shows the same isochrone, offset by $+1.0$ mag in $M_H$ for clarity purposes,
  while the blue plus signs represent points along that isochrone for a regular
  grid of initial masses ($\Delta M_{\rm i} = 0.08$ in this case) to represent a
  uniform mass sampling cadence equivalent to a full IMF sampling.
  Note the two bright red supergiants in cluster \#111, with only one
  counterpart at the blue end of the blue loop.   
\emph{Panel (b)}: the black line represents the integrated spectrum of
  cluster \#111, normalized to unity at $\lambda = 1 \mu$m. The red line
  represents the co-added spectra of the two red supergiants shown in panel
  (a), using the same normalization factor. 
For comparison, the purple line represents the spectrum of the blue supergiant
in cluster \#111. Finally, the blue and grey lines represent spectra for
  SSPs with (log\,(age), [Z/H]) = (7.0, $-$0.4) and (10.1, 0.0), respectively,
normalized to fit the flux of cluster\#111 at $\lambda = 2.2 \mu$m.
Note that the NIR spectrum of cluster \#111 is dominated by its red
  supergiants, and fit much better by a log\,(age) = 10.1 SSP spectrum than by that
  of its own log\,(age) = 7.0.} 
\label{f:ageproblem_7p0}
\end{figure}

An interesting example of this is illustrated in
Figure~\ref{f:ageproblem_7p0}. It shows a $M_H$ vs.\ log\,$T_{\rm
    eff}$ diagram\footnote{We chose log\,$T_{\rm eff}$ for the abscissa to
    maximize the resolution of the stars' sampling of the isochrone.} of a
simulated $10^4\: M_{\odot}$ cluster with an age of 10 Myr (and 
[Z/H] = $-$0.4), which happens to have two supergiants at the red (cool) end of
the blue loop and one at its blue (hot) end. The lower panel of
Figure~\ref{f:ageproblem_7p0} shows that the NIR SED of that cluster is
completely dominated by its two red supergiants, which causes the SED of that
cluster to be best fit by an SSP with (log\,(age), [Z/H]) = (10.1, 0.0), thus
masquerading as a SSP three orders of magnitude older (and hence also much more
massive) than the actual cluster. In strong contrast, the age of this
  same cluster was recovered correctly to within 0.1 dex in the 
  blue optical regime. 

\begin{figure*}
\centerline{
\includegraphics[width=0.48\textwidth]{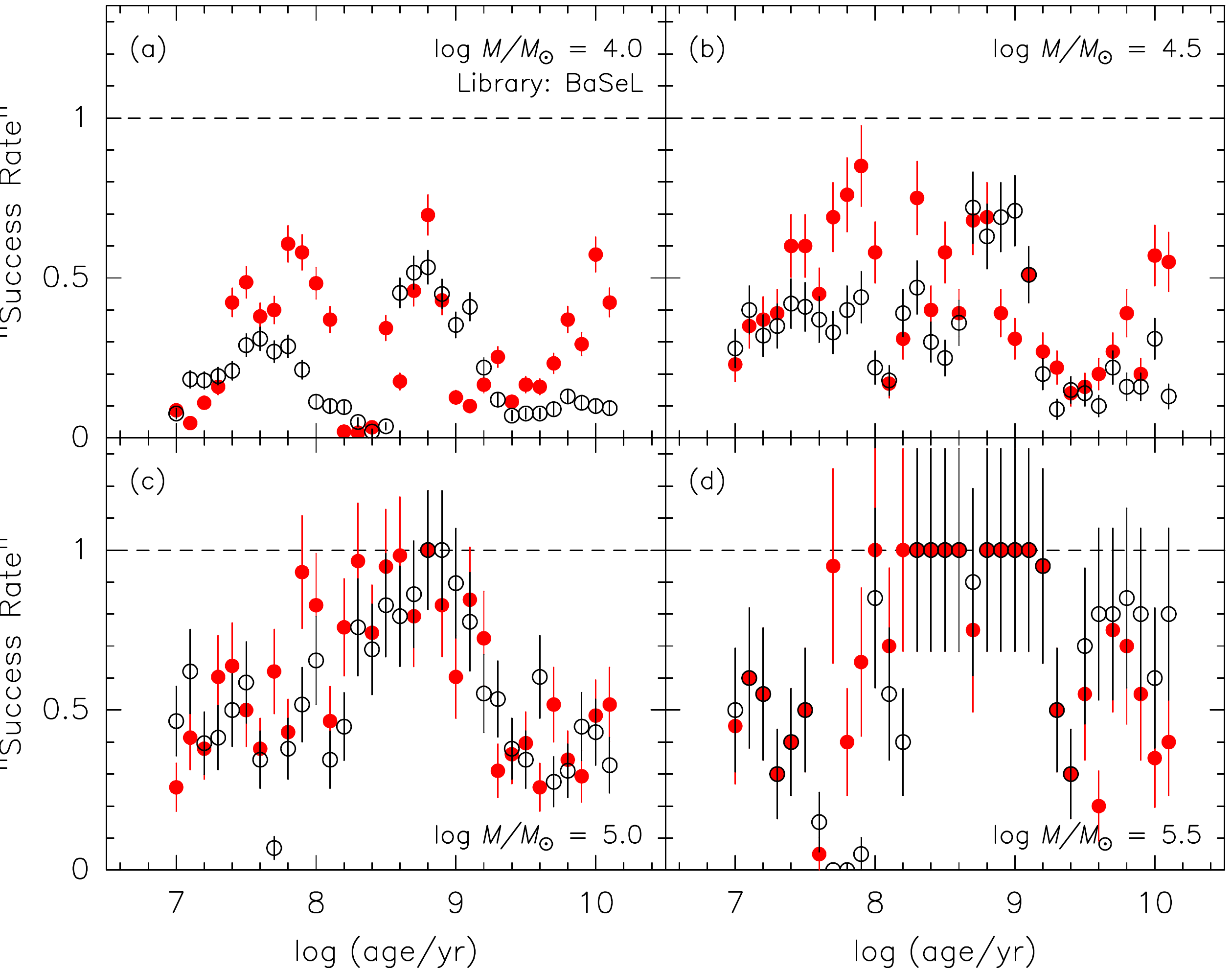}
\hspace*{1mm}
\includegraphics[width=0.48\textwidth]{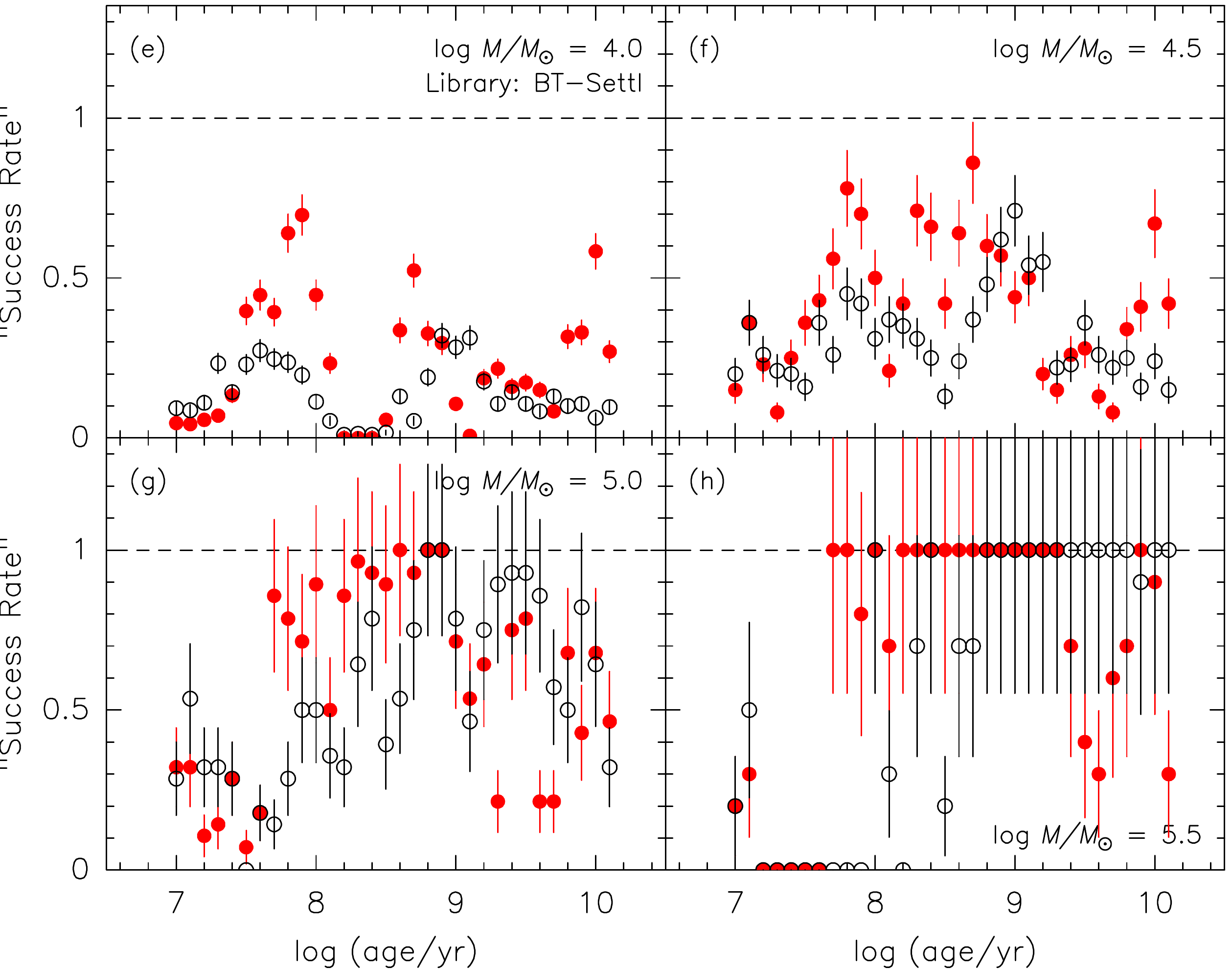}
}
\caption{Same as Figure~\ref{f:fractions_age_ZH_opt}, but now for the NIR range.}
\label{f:fractions_age_ZH_NIR}
\end{figure*}

The success rate of age determination increases significantly in the age range
of 30\,--\,100 Myr, during which increasingly larger percentages of stars
populate the ``red'' branches of shell helium (or hydrogen) burning stars (see
Figure~\ref{f:cmds_NIR}). This success rate then again decreases strongly for
low-mass clusters at an age of $\sim$\,150 Myr, when the AGB suddenly starts to
extend far beyond the blue loop in NIR luminosity, corresponding to the onset of
the TP-AGB phase\footnote{This phase in stellar evolution is sometimes referred
  to as the ``AGB phase transition'' \citep[see, e.g.,][]{RB86,Maraston05}.}
\citep[e.g.,][]{Marigo08}. 
 As the luminous TP-AGB is relatively sparsely populated in this
  age range, stochastic fluctuations have a strong impact. Among 
  low-mass clusters, this causes actual discontinuities in the distribution of 
  integrated NIR and mid-IR fluxes (see panel (d) in
  Figure~\ref{f:specplot_all}).    
However, this trend disappears for higher-mass clusters (see
Figure~\ref{f:fractions_age_ZH_NIR}), due to their more uniform
sampling of the extent of the TP-AGB.   

After an age of $\sim$\,600 Myr, the population of stars on the TP-AGB increases
with increasing age, which causes the success rate of age determination to
increase as well (mainly for the lower-mass clusters). This trend stops at an
age of $\sim$\,1 Gyr, when the RGB phase transition occurs and the NIR SED
starts to depend critically on the distribution of stars along the AGB and
RGB.
 As an example, for a low cluster mass of $10^4\: M_{\odot}$ and
  log\,(age/yr) = 9.1, our results show that the particular sampling of the RGB
  and AGB can cause the best-fit SSP log\,(age) to vary widely from 7.1 to 9.9
  in the NIR regime. 
  Panel (e) in Figure~\ref{f:specplot_all} also illustrates the strong effect of
  stochastic fluctuations around this age. 

After the RGB phase transition era, the success rate of age determination
gradually increases again with increasing age, due to the increasingly uniform
sampling of the RGB and AGB. 

\begin{figure}
\centerline{\includegraphics[width=8.cm]{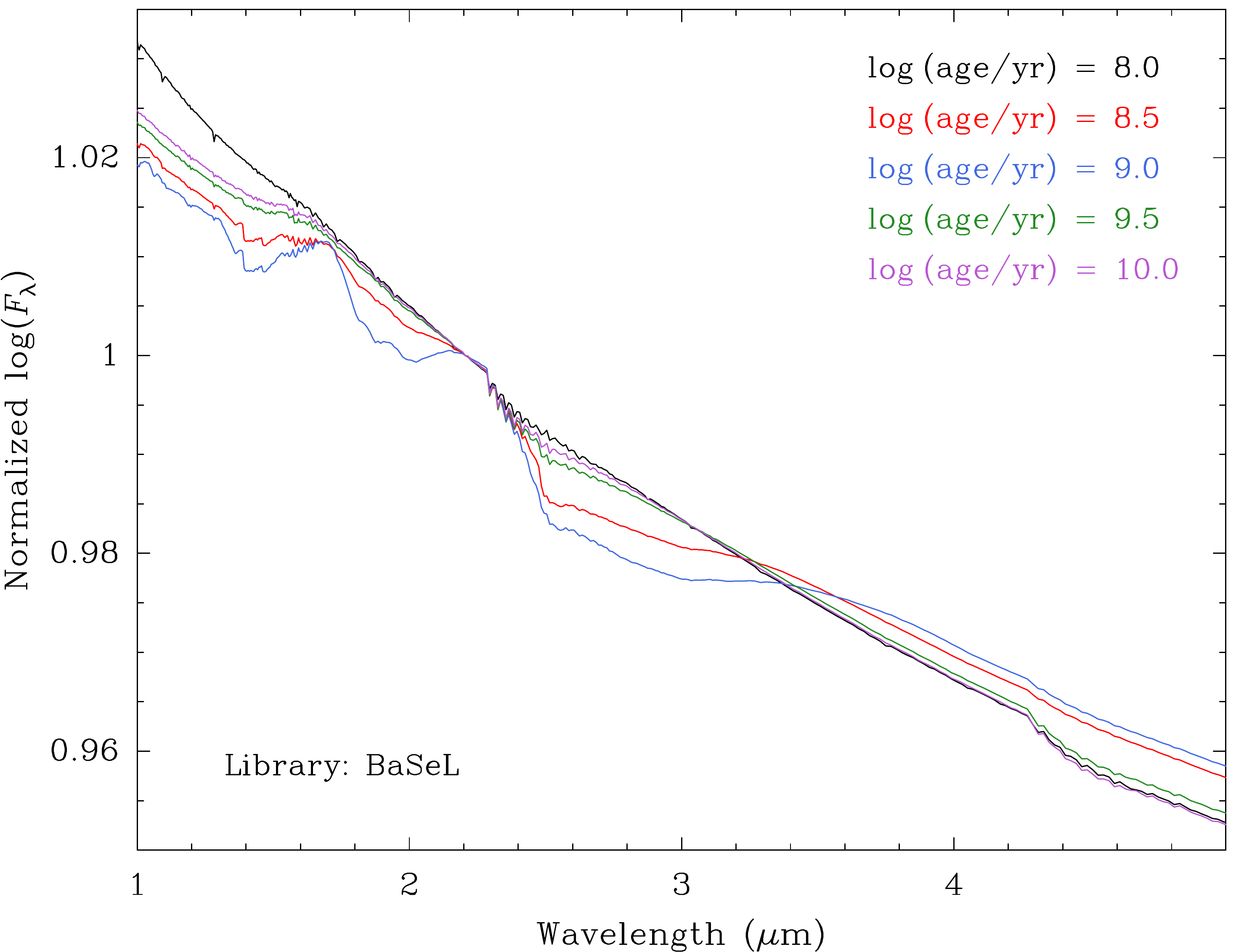}}
\caption{1\,--\,5 $\mu$m SEDs of synthetic SSPs with [Z/H] = $-$0.4. Ages and
  associated line colours are shown in the legend. The SEDs are normalized to
  unity at 2.2 $\mu$m. Note the strong H$_2$O bands at $\sim$\,1.4, 1.9, and 2.8
  $\mu$m along with an excess of continuum radiation beyond the 2.8 $\mu$m band
  for the SSPs with log\,(age/yr) = 8.5 and 9.0, when the contribution of cool
  TP-AGB stars is largest.} 
\label{f:midIR_SEDs}
\end{figure}

In the NIR range, the success rate of metallicity determination as functions of
age and cluster mass follows that of the success rate of age determination (see
Figures~\ref{f:fractions_age_ZH_NIR}, \ref{f:offsets_age_ZH_NIR}, and
\ref{f:offsets_age_ZH_NIR_BT}). This is largely due to the strong and wide
molecular bands of H$_2$O at $\sim$\,1.4 and 1.9 $\mu$m which are only produced
by the coolest TP-AGB stars that provide a significant fraction of the NIR flux
in the age range of $8.5 \la \mbox{log\,(age/yr)} \la 9.2$. This is illustrated
in Figure~\ref{f:midIR_SEDs} and  further discussed in the next Section. 

Finally, Figure~\ref{f:fractions_age_ZH_NIR} shows that BaSeL and BT-Settl
libraries generally yield similar success rates of age and metallicity
determination in the NIR regime. The only qualitative difference between the two
is that the higher spectral resolution of the BT-Settl library provides somewhat
better metallicity determination for massive clusters with older ages ($\ga$\,2
Gyr).  

\subsection{Mid-IR Range: 2.5\,--\,5.0 $\mu$m}
\label{s:MIR}

In the mid-IR regime, the post-MS stages of stellar evolution contribute the
highest luminosity fraction in this regime relative to those at shorter
wavelengths (see Figure~\ref{f:cmds_midIR}). As such, stochastic fluctuations
have a large impact on the \emph{luminosities} of low-mass star clusters in the
mid-IR regime. However, the spectral radiance of electromagnetic radiation in
the mid-IR is well approximated by the Rayleigh-Jeans law, and as such, the
shapes of SEDs of SSPs of different ages, as well as SEDs of different clusters
of a given mass,  generally look more similar to one another than at shorter
wavelengths (see Figure~\ref{f:specplot_all}).  

The shape of isochrone CMDs in the mid-IR, shown in Figure~\ref{f:cmds_midIR},
emphasizes this in that differences in $K-M$ colour between the various phases
of stellar evolution stay within $\sim$\,0.3 mag. The overall similarity of
shapes of mid-IR CMDs of isochrones of young ($\la$ 300 Myr) and old ($\ga$ 3
Gyr) SSPs is especially remarkable.  

However, in the age range considered here, there is one phase of stellar
evolution with a recognizable impact to shapes of mid-IR SEDs. This is the
TP-AGB, which dominates the mid-IR luminosity in the age range of $8.5 \la
\mbox{log\,(age/yr)} \la 9.2$. It extends to effective temperatures that are
even lower than the low-mass end of the MS. In our synthetic SSP SEDs, the main
spectral features caused by the cool TP-AGB stars are (1) wide and relatively
strong H$_2$O bands at $\sim$\,1.4, 1.9, and 2.8 $\mu$m\footnote{We recall
  that we model TP-AGB stars as oxygen-rich M stars. However, if
  instead the TP-AGB happens to be dominated by carbon stars in a given cluster
  in this age range, then the SED will have equally recognizable features from
  CN, C$_2$, HCN, and C$_2$H$_2$ in the 1\,--\,5 $\mu$m region \citep{Aringer09}.},
and (2) an excess of continuum radiation beyond the $\sim$\,2.8 $\mu$m band of
H$_2$O relative to SSPs with younger or older ages (see Figure~\ref{f:midIR_SEDs}).
With this in mind, we use $\lambda_{\rm norm}$ = 3.5 $\mu$m for full-spectrum 
fitting in the mid-IR regime.

\begin{figure*}
\centerline{
\includegraphics[width=0.48\textwidth]{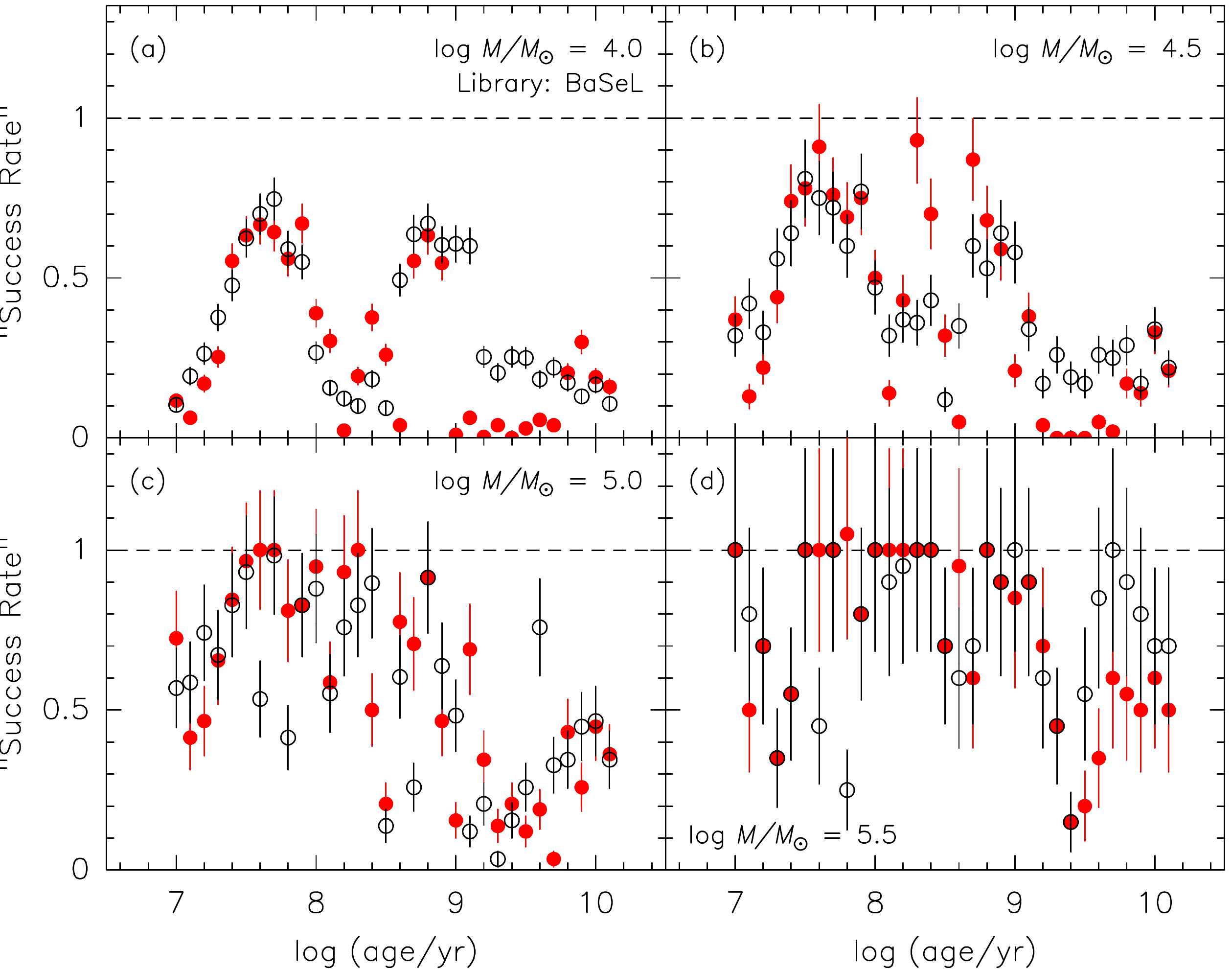}
\hspace*{1mm}
\includegraphics[width=0.48\textwidth]{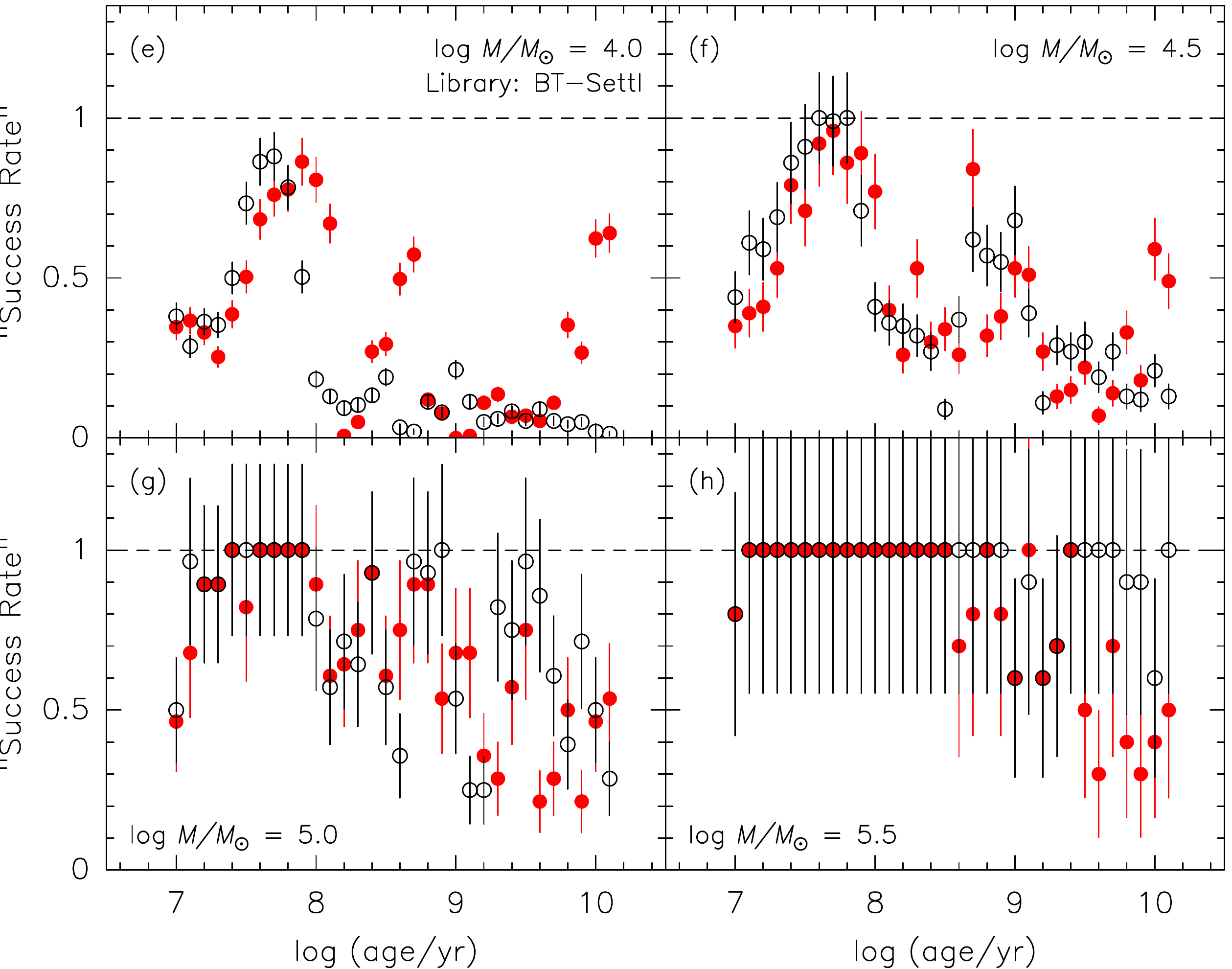}
}
\caption{Same as Figure~\ref{f:fractions_age_ZH_opt}, but now for the mid-IR range.}
\label{f:fractions_age_ZH_midIR}
\end{figure*}

The H$_2$O bands mentioned above are intrinsically strong, especially the 2.8
$\mu$m band, which is the strongest spectral absorption feature in the
2.5\,--\,5 $\mu$m mid-IR range when the physical environment is conducive to the
formation of H$_2$O bands. And indeed, the success rate of metallicity
determination using full-spectrum fitting in the mid-IR as functions of age and
cluster mass follows that of the success rate of age determination. In fact, it
does so even more closely than in the NIR (see
Figure~\ref{f:fractions_age_ZH_midIR}, and Figures~\ref{f:offsets_age_ZH_midIR}
and  \ref{f:offsets_age_ZH_midIR_BT} in the Appendix). This is likely due to the
fundamental (1-0) CO bandhead at 4.6 $\mu$m, which is stronger than the first
overtone CO bandhead at 2.3 $\mu$m.  

\begin{figure*}
\centerline{
\includegraphics[width=0.4\textwidth]{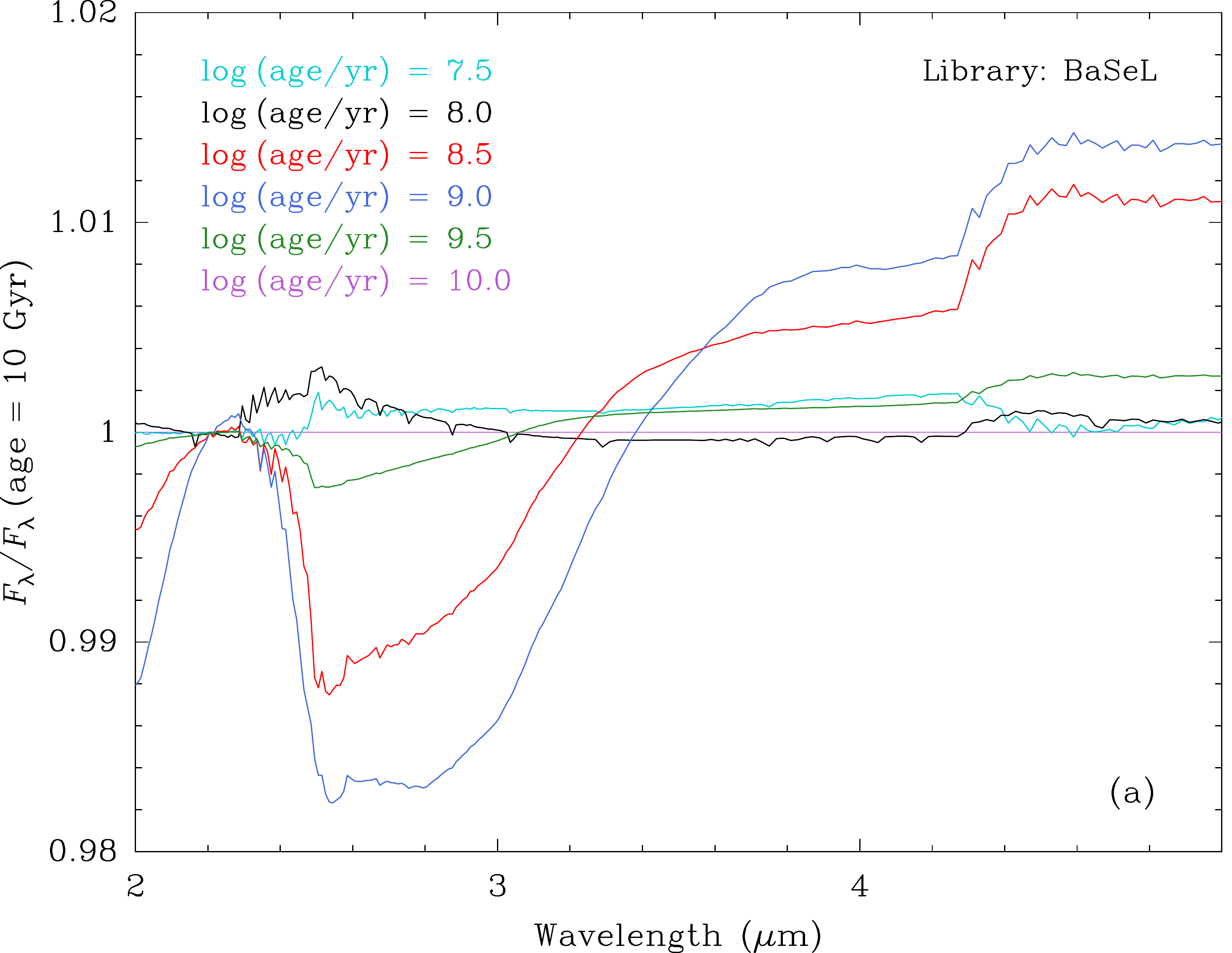}
\hspace*{1mm}
\includegraphics[width=0.4\textwidth]{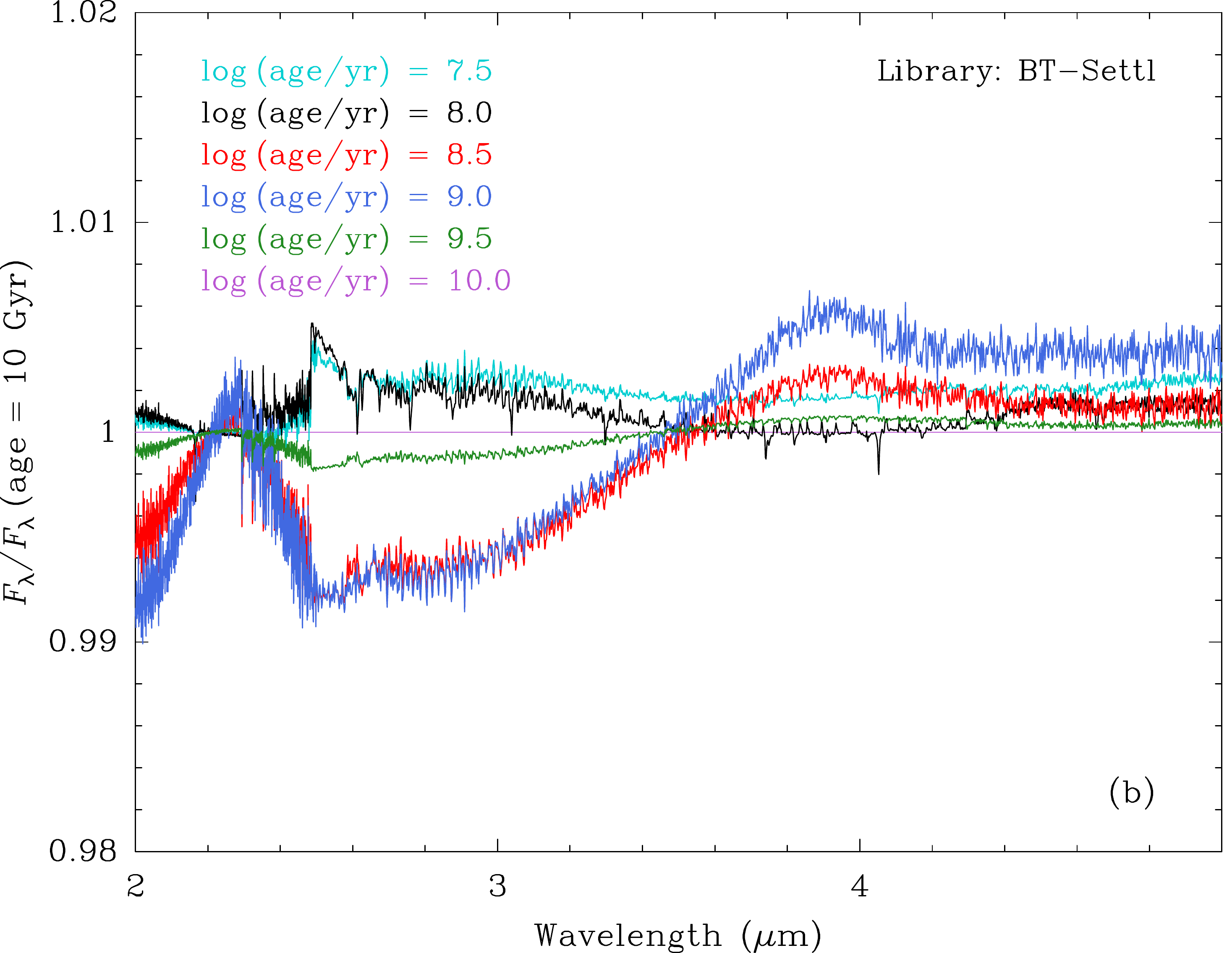}
}
\caption{2\,--\,5\,$\mu$m flux ratios of SEDs of SSPs relative to that of a SSP
  with an age of 10 Gyr. The SED curves are normalized to unity at 2.2 $\mu$m to
  highlight shape differences. Different line colours correspond to different
  ages as shown in the legend. All SSPs shown have [Z/H] = $-$0.4. Panel (a)
  shows the ratios for the BaSeL spectral library while panel (b) does so for
  the BT-Settl library. See discussion in Section~\ref{s:MIR}.} 
\label{f:midIRcomp}
\end{figure*}

\begin{figure}
\centerline{
\hspace*{1mm}
\includegraphics[width=6.5cm]{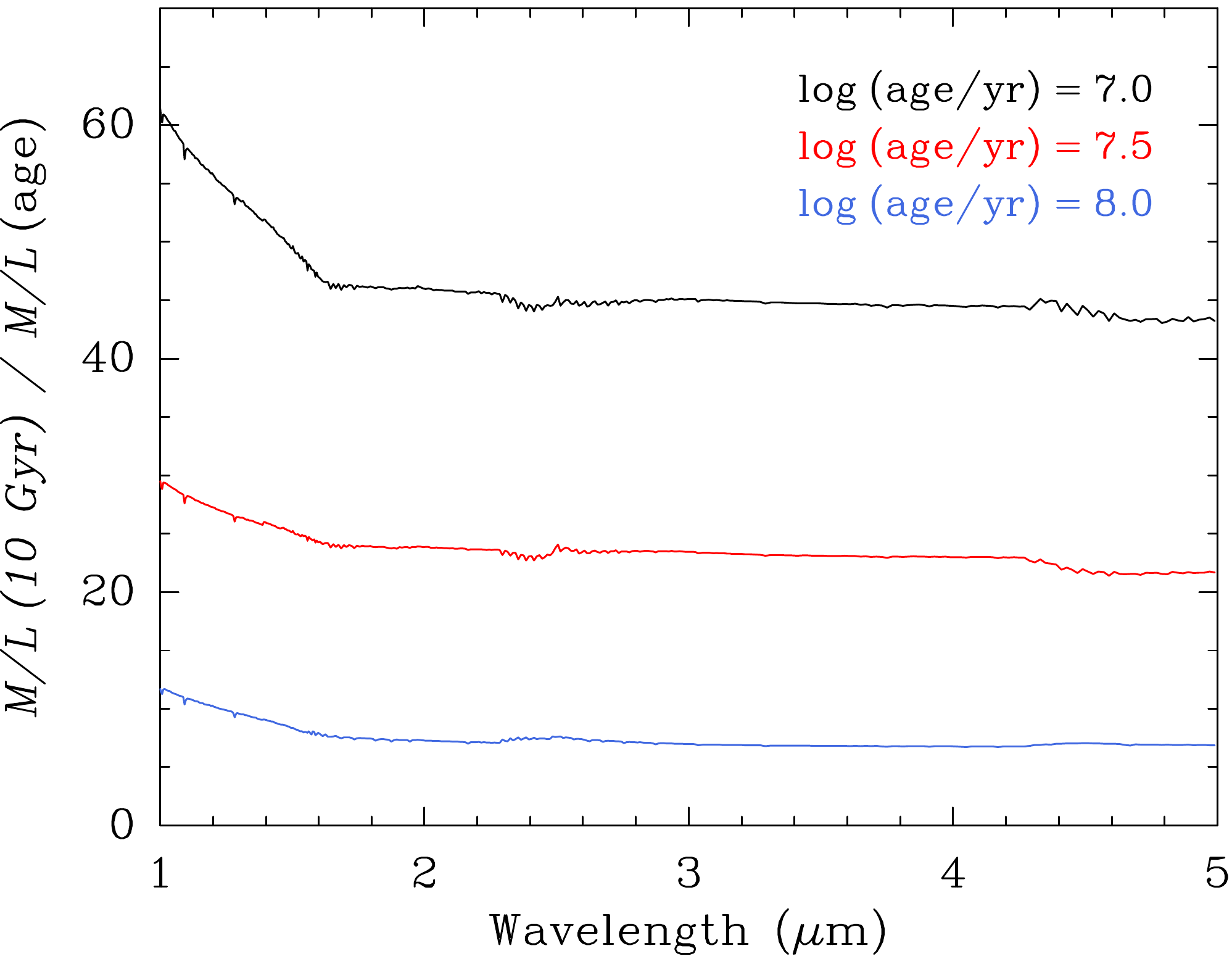}
}
\caption{$M/L$ in the 1\,--\,5 $\mu$m region of a 10-Gyr-old SSP relative to
  those of SSPs of ages 10 Myr, 30 Myr, and 100 Myr (see legend), using the
  BaSeL library. See discussion in Section~\ref{s:MIR}.} 
\label{f:MLplot}
\end{figure}

In terms of the impact of different spectral resolutions,
Figure~\ref{f:fractions_age_ZH_midIR} shows that the use of the BaSeL
vs.\ BT-Settl libraries does not translate in any significant differences in the
success rate of age and metallicity determination (as defined in
Section~\ref{s:optical}) in the mid-IR regime. However, a comparison of
Figures~\ref{f:offsets_age_ZH_midIR} and \ref{f:offsets_age_ZH_midIR_BT} shows
that the higher spectral resolution of the BT-Settl library does translate in a
significant improvement in terms of the accuracy and precision of age
determination at ages $\ga$\,2 Gyr. Specifically, the use of the BaSeL library
causes a bimodal distribution of the fitted age for those ages, which does not
occur for the BT-Settl library. To understand this difference, we compare SSP
SEDs in the 2\,--\,5 $\mu$m region after normalizing by the SED for an age of 10
Gyr in Figure~\ref{f:midIRcomp}. It can be seen that the aforementioned
difference is due, at least in part, to narrow spectral features such as the
Brackett and Pfund series of Hydrogen that occur in hot stars in young SSPs,
which are smoothed over by the low resolution of the BaSeL library. Note however
that these intrinsic differences between ``young'' and ``old'' SSP SEDs in the
mid-IR regime are only at the \% level. Hence, detection of these differences
in observed spectra will require relatively high signal-to-noise (S/N)
ratios. On the other hand, the $M/L$ ratio in the mid-IR regime for a SSP with
an age of 10 Gyr is factors $\sim$\,7\,--\,45 higher than for SSPs with
ages between 10 and 100 Myr (see Figure~\ref{f:MLplot}), thus rendering
a detection of a luminous cluster in the mid-IR to be much more likely to be
young rather than old.  

\subsection{The cluster mass dependence of age and [Z/H] recovery in
  different wavelength regimes} 
\label{s:massdep}

\begin{figure*}
\centerline{\includegraphics[width=16.cm]{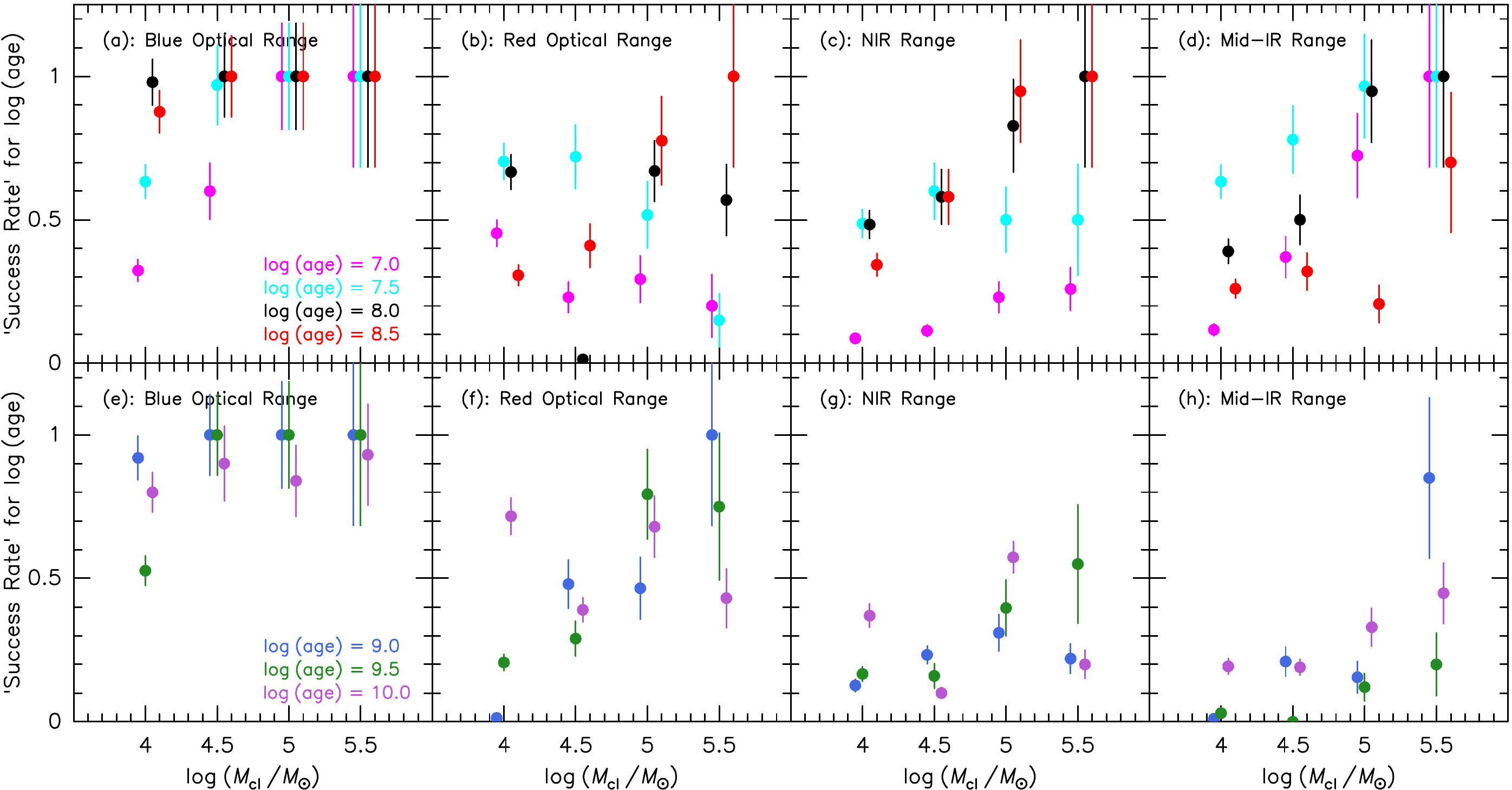}}
\caption{Success rate of age determination as a function of cluster mass $M_{\rm
    cl}$, for the four wavelength regimes discussed in this paper: blue optical
  range (panels a and e), red optical range (panels b and f), NIR range (panels
  c and g), and mid-IR (panels d and h). Different ages are indicated by
  different symbol colours (see legend in panels a and e). To improve clarity,
  the symbols for different ages within a given panel are slightly offset in
  log\,($M_{\rm cl}$) from one another (by $\Delta \log\,(M_{\rm cl})$ = 0.05).}  
\label{f:massdep_age_success}
\end{figure*}

\begin{figure*}
\centerline{\includegraphics[width=16.cm]{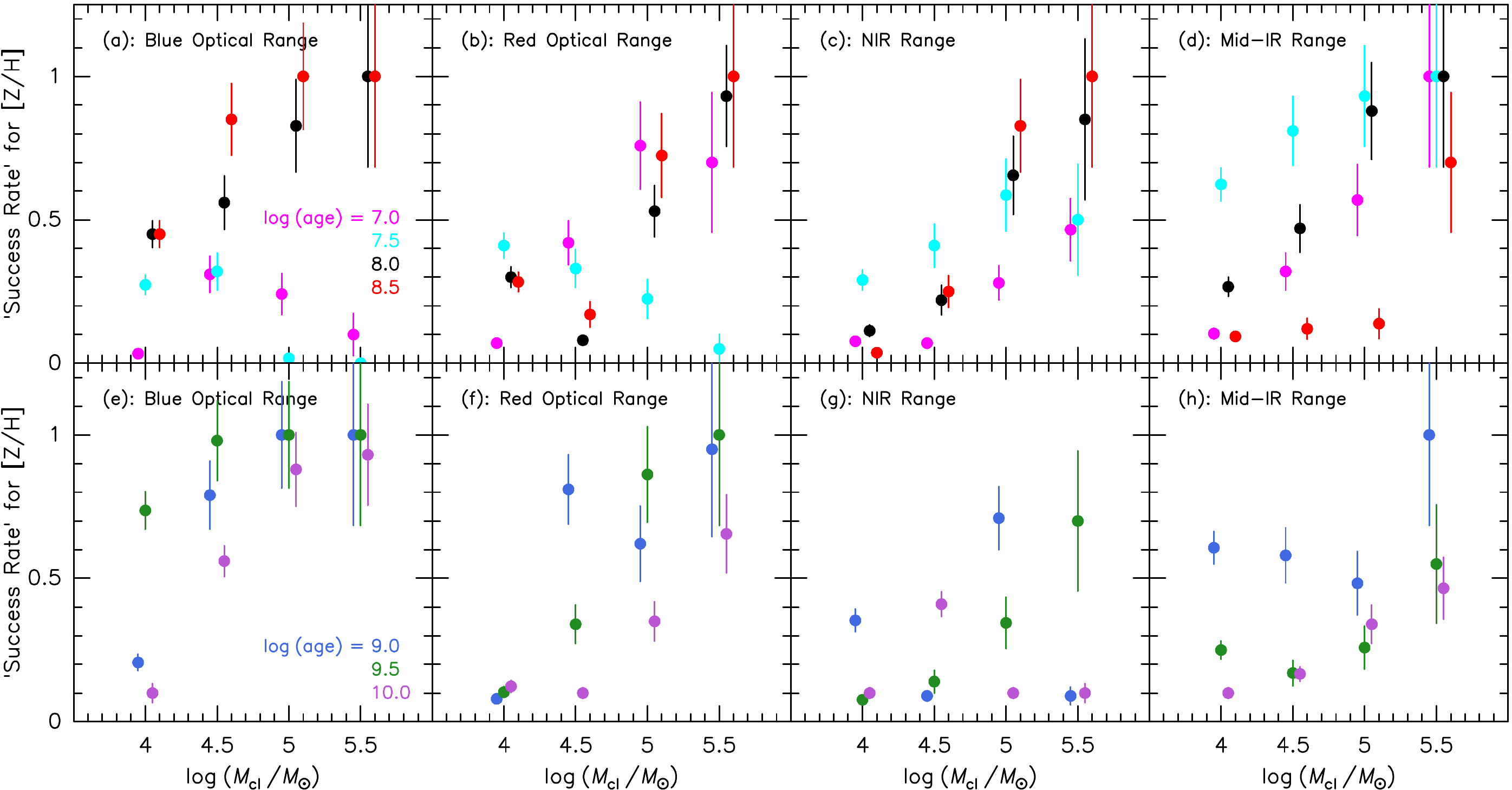}}
\caption{Similar to  Figure~\ref{f:massdep_age_success}, but now plotting the success
  rate of [Z/H] determination. Symbols are the same as in
  Figure~\ref{f:massdep_age_success}.}  
\label{f:massdep_ZH_success}
\end{figure*}

\begin{figure*}
\centerline{\includegraphics[width=16.cm]{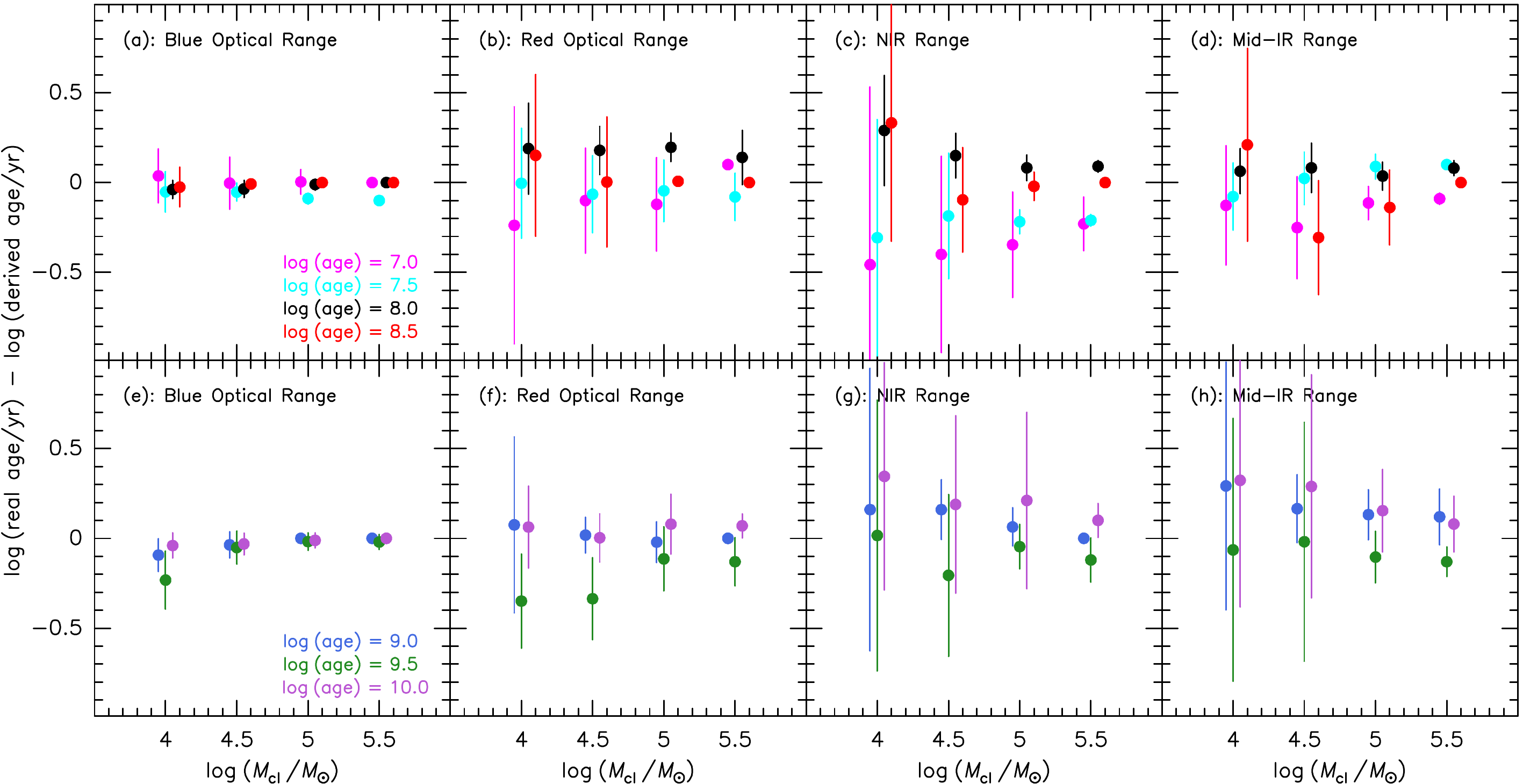}}
\caption{Similar to  Figure~\ref{f:massdep_age_success}, but now plotting the mean
  offset of the best-fit log\,(age) from the true value along with its standard
  deviation among the simulated clusters. Symbols are the same as in
  Figure~\ref{f:massdep_age_success}.}  
\label{f:massdep_age}
\end{figure*}

\begin{figure*}
\centerline{\includegraphics[width=16.cm]{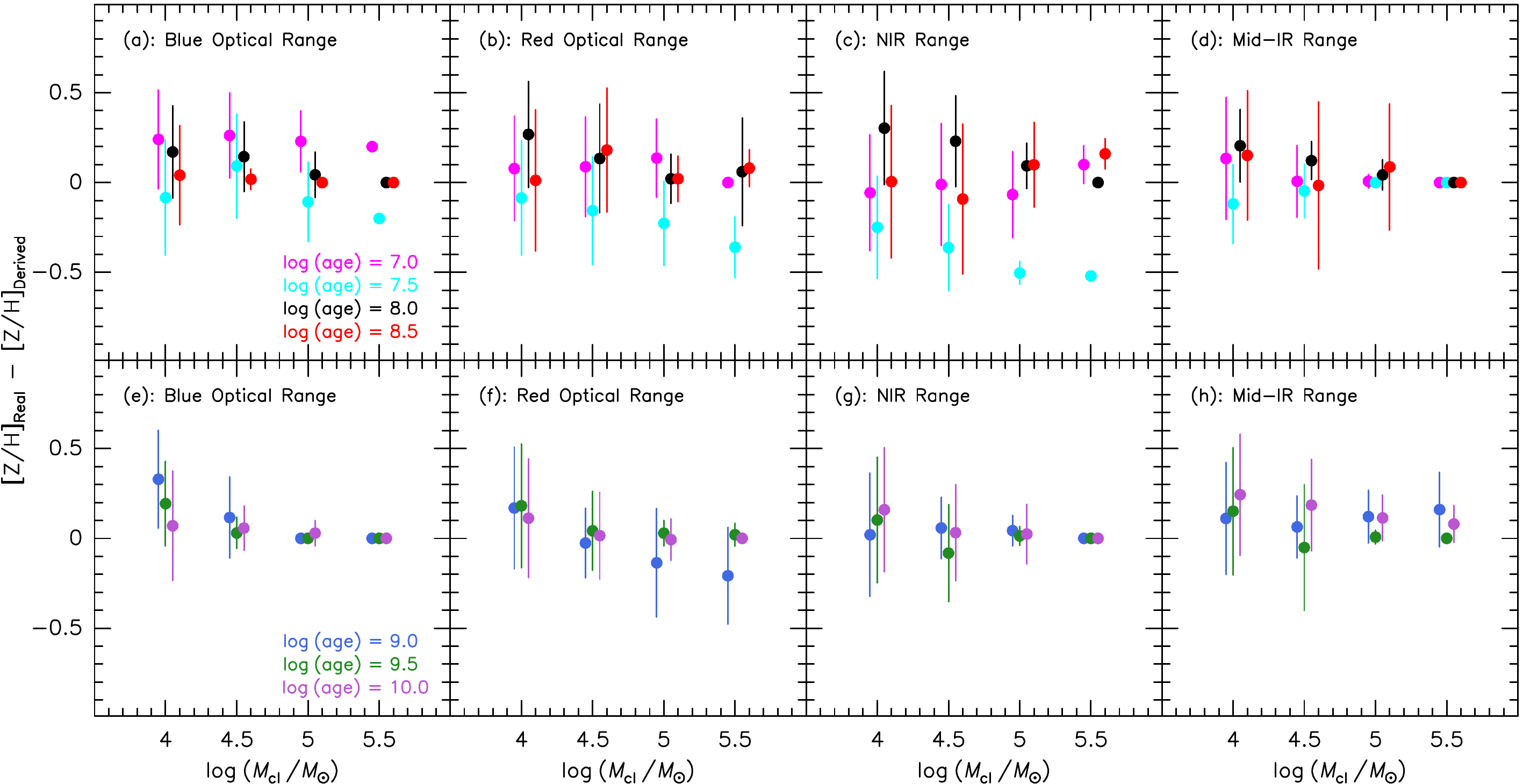}}
\caption{Similar to  Figure~\ref{f:massdep_age_success}, but now plotting the mean
  offset of the best-fit [Z/H] from the true value along with its standard
  deviation among the simulated clusters. 
  Symbols are the same as in Figure~\ref{f:massdep_age_success}.}  
\label{f:massdep_ZH}
\end{figure*}

\begin{figure}
\centerline{\includegraphics[width=7.cm]{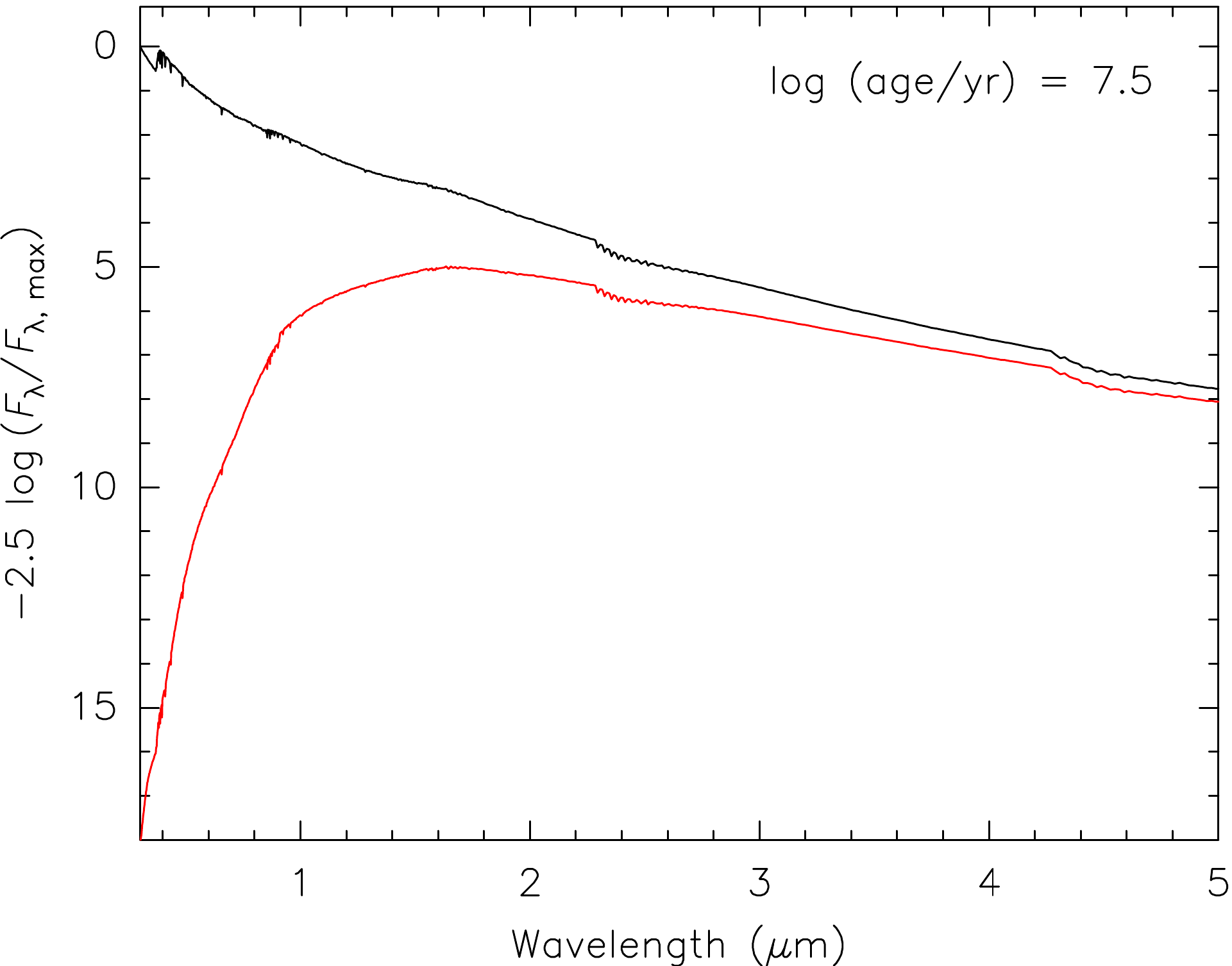}}
\caption{The black line represents the 0.3\,--\,5 $\mu$m SED of a SSP with
  log\,(age/yr) = 7.5 in magnitude flux units, normalized to 0 at the wavelength
  of maximum flux. The red line represents the same SED after applying $A_V$ =
  10 mag of extinction using the \citet{Cardelli89} prescription for $R_V$ = 3.1. }   
\label{f:reddening}
\end{figure}

Using the results described above in Sections~\ref{s:optical}\,--\,\ref{s:MIR},
we illustrate the cluster mass dependence of the success rate of age and
metallicity determination in the various wavelength regimes in
Figures~\ref{f:massdep_age_success} and \ref{f:massdep_ZH_success}, 
 while Figures~\ref{f:massdep_age} and \ref{f:massdep_ZH} show the cluster
  mass dependence of the mean offset and standard deviation of derived
  log\,(age) and [Z/H] from the true values. 
To avoid overcrowding the plots, the results are only shown for
$7.0 \leq \mbox{log\,(age/yr)} \leq 10.0$ with a step size of 0.5 dex.
These Figures allow a quick comparison of the power 
of the various wavelength regimes, not just in terms of the cluster mass
dependence of the accuracy and precision of age and metallicity determination,
but also in terms of the identification of certain types of populations through
full-spectrum fitting. For example, if the goal is to identify young massive
clusters (YMCs, with ages $\la$ 100 Myr and masses $\ga$ 10$^5 \: M_{\odot}$) in
distant galaxies,
Figures~\ref{f:massdep_age_success}\,--\,\ref{f:massdep_ZH} show that
while the blue optical range is in principle best suited for that goal, the
mid-IR range is a good alternative (and actually better than the blue optical
range in terms of [Z/H] measurement precision, due to the strong CO (1-0)
bandhead at 4.6 $\mu$m). Moreover, star forming regions typically exhibit
 high and non-uniform extinction, and extinction in the mid-IR is factors
$\ga$ 10$^4$ lower than in the blue optical range, thus allowing surveys
that are significantly less impacted by regions of high extinction. This is
  illustrated in Figure~\ref{f:reddening} for an SSP with log\,(age/yr) = 7.5
  and an extinction $A_V$ = 10 mag, which is not untypical for YMCs detected in
  optical images of star-forming galaxies \citep[e.g.,][]{Wilson06}. Finally,
we note that the omission of the influence of circumstellar dust around upper
AGB stars in our modeling (cf.\ Section~\ref{s:mockGCs}.2) is not expected to
have an impact on this result, since AGB stars with circumstellar dust shells
only start having a significant contribution to the integrated mid-IR light at
an age of about 200 Myr, after the YMC era \citep[see, e.g.,][]{Villaume15}. 

\section{Summary}
\label{s:summ}

We studied the effects of stochastic fluctuations of stellar masses near the top
of the stellar mass function in star clusters as a function of the star cluster
mass in terms of variations of SSP-equivalent ages and metallicities derived
using the technique of full-spectrum fitting, and how the level of those
variations depends on star cluster mass (in the range of $10^4 \leq M/M_{\odot}
< 10^6$) and wavelength (in the range of $0.3 \leq \lambda/\mu\mbox{m} \leq 5$).  
To this end, we created a large set of simulated star cluster spectra which we
compared with simulated ``true'' SSP spectra, using the same isochrones and
spectral libraries. The age range considered in our study is $7.0 \leq
\mbox{log\,(age/yr)} \leq 10.1$. We also compared the results when using two
different spectral libraries (BaSeL and BT-Settl), featuring different physical
model ingredients and spectral resolutions.  
We analyzed four different spectral regimes: blue optical (0.35\,--\,0.7
$\mu$m), red optical (0.6\,--\,1.0 $\mu$m), near-IR (1.0\,--\,2.5 $\mu$m), and
mid-IR (2.5\,--\,5.0 $\mu$m). The results can be summarized as
follows. \\ [-4ex]  
\begin{itemize}
    \item In general, the blue optical regime (0.35\,--\,0.7\,$\mu$m) is by far
      the least impacted by stochastic fluctuations among the wavelength regimes
      studied here. This is reflected in it yielding the highest accuracy and
      precision of age determination (for clusters of all ages) and metallicity
      determination (for ages $\ga$ 300 Myr). With the exception of the youngest
      clusters in our simulations (log\,(age/yr) $\la$ 7.5), the full precision
      of age and metallicity determination through full-spectrum fitting in the
      blue optical regime is already reached at a cluster mass of $\sim 3 \times
      10^4 \: M_{\odot}$, which is a factor $\ga$\,10 lower than for all other
      wavelength regimes (with $\lambda > 0.6 \: \mu$m) studied here. However,
      metallicity determination for SSPs with log\,(age/yr) $\la$ 8.5 suffers
      from significant systematic errors (of order $\la$\,0.5 dex) in this
      regime. These findings are similar for the BaSeL and BT-Settl spectral
      libraries. 
    \item The red optical regime (0.6\,--\,1.0\,$\mu$m) and the near-IR regime
      (1.0\,--\,2.5\,$\mu$m) have similar power in terms of their accuracy and
      precision of age and metallicity determination using full-spectrum
      fitting. The impact of stochastic fluctuations is large across virtually
      all ages in these wavelength regimes, leading to errors that can reach
      beyond 2 dex in age and beyond 0.5 dex in [Z/H] for clusters with
      $M/M_{\odot} < 10^5$, depending on the age. The largest cluster mass
      dependence of the precision of age determination is found around the ages
      where the luminous TP-AGB or RGB stages start to be populated (i.e.,
      log\,(age/yr) $\sim$\,8.4 and 9.0, respectively), especially in the NIR
      regime. The choice of spectral library (BaSeL vs.\ BT-Settl) has little
      impact on our findings in the red optical and NIR regimes. 
    \item Stochastic fluctuations also have a large impact on the luminosities
      of star clusters in the mid-IR regime (2.5\,--\,5.0\,$\mu$m). This is
      because the post-MS stages of stellar evolution contribute the highest
      luminosity fraction in this regime relative to those at shorter
      wavelengths. However, with the important exception of clusters older than
      $\approx$\,2 Gyr (see below), these fluctuations in luminosity do not
      generally cause significant inaccuracy of age determination in the mid-IR
      regime. For the age range in which the TP-AGB provides a significant
      contribution to the luminosity, this is caused by two features that
      clearly identify this stage: a strong H$_2$O band at 2.8 $\mu$m and a
      significant excess of continuum radiation at $\lambda \ga 3 \: \mu$m due
      to the cool TP-AGB stars. However, the mid-IR SEDs of SSPs with age
      $\ga$\,2 Gyr are generally very similar to those with age $\la$\,100
      Myr. Use of the low-resolution BaSeL spectral library causes this to yield
      an approximately bimodal distribution of best-fit ages for old clusters,
      especially for those with $M/M_{\odot} \leq 10^5$. This bimodality
      disappears when using the higher-resolution BT-Settl library, and it also
      does not occur for young clusters in our tests.   
    
      In fact, we find the mid-IR to be a powerful regime for the purpose of
      surveying and identifying young massive clusters in star-forming
      regions. While its accuracy and precision of age determination for ages
      $\la$\,600 Myr is slightly below that of the blue optical region, the mid-IR
      is significantly less impacted by the strong extinction that is common in
      star-forming regions. 
\end{itemize}

\section*{Acknowledgments}
We thank Morgan Fouesneau for providing us access to the high-resolution
BT-Settl spectral library as an add-on to his {\tt pystellibs} GitHub
repository. RA thanks the Space Telescope Science Institute for a sabbatical
visitorship including travel and subsistence support as well as access to their
science cluster computer facilities. 
We acknowledge use of the software packages NumPy \citep{NumPy}, AstroPy
\citep{astropy:2013, astropy:2018}, Matplotlib \citep{Hunter07}, and Gildas
\url{(https://www.iram.fr/IRAMFR/GILDAS)}.  

\section*{Data availability}
The model spectra of all mock star clusters and SSPs created for this
paper are available through the
\href{http://www.stsci.edu/~goudfroo/data_GA20}{website of the
  corresponding author}. They will also be made available through
\href{http://cdsarc.u-strasbg.fr/viz-bin/qcat?VI/157}{CDS} once this
paper is published.  

\bibliographystyle{mnras}
\bibliography{References}

\appendix

\section{Plots of the distribution of SSP fitting results}
\label{s:distr_plots}

In this appendix, we show plots of the full  distributions the
differences of the ages and metallicities of all simulated clusters
found through full-spectrum fitting with respect to the "true" input
values. These distributions are shown as a function of input age, and
separately for each wavelength regime discussed in this paper.  

\begin{figure*}
\centerline{\includegraphics[width=16.cm]{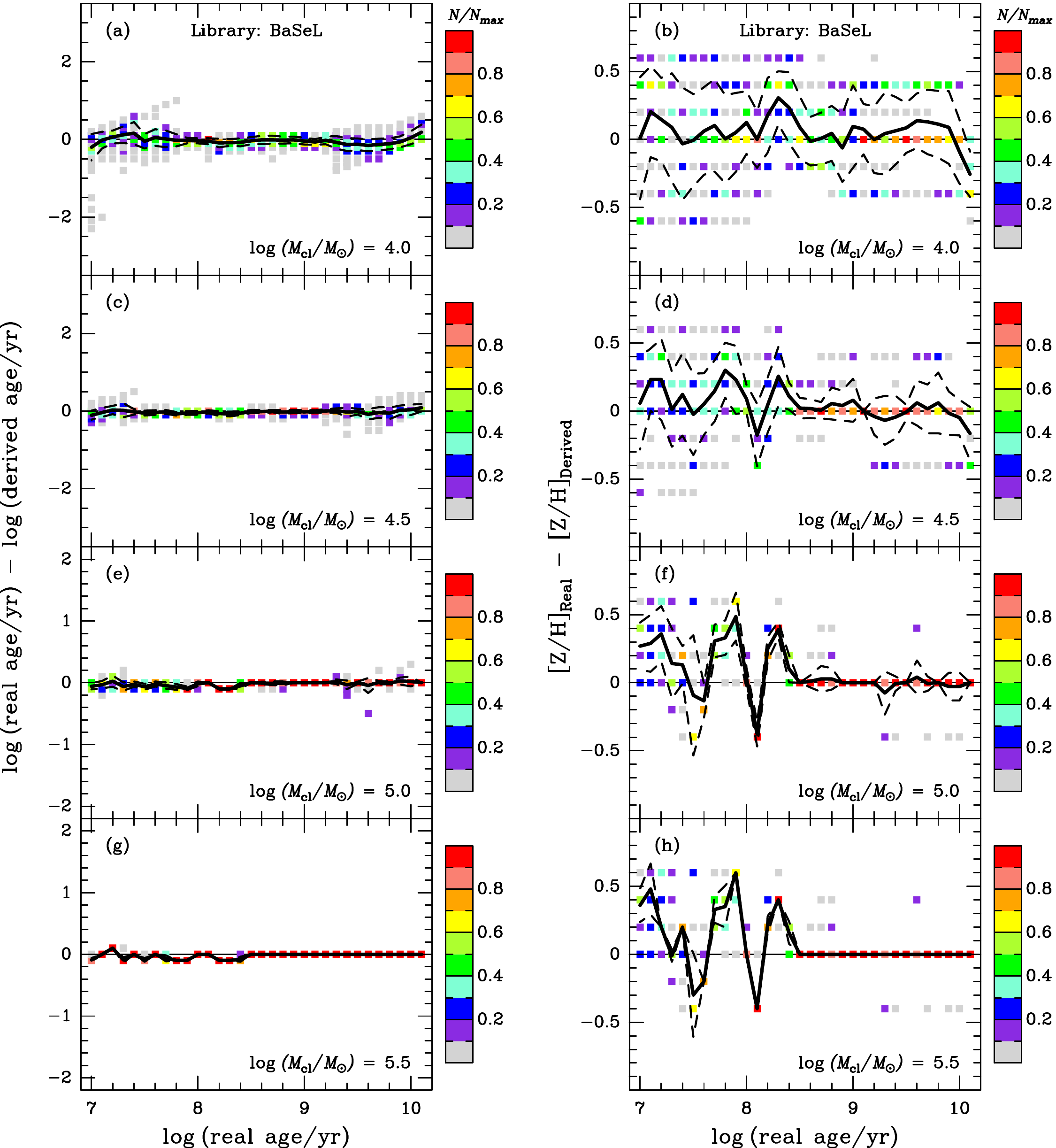}}
\caption{Blue optical range:\ SSP fitting results for age (left
  panels) and [Z/H] (right panels) expressed as offsets of the best
  fit in log\,(age) or [Z/H] from the true value, as a function of
  log\,(age). The symbol colour indicates the relative number of times
  that offset occurred in the results (see colour bar on the right
  side of each panel). Different rows of panels show the results for
  different cluster masses $M_{\rm cl}$ (see legends). The thick black
  solid line in each panel indicates the average offset, and the two
  dashed lines indicate $\pm 1 \sigma$ from the average offset. The
  BaSeL spectral library is used for the results in this Figure.} 
\label{f:offsets_age_ZH_opt}
\end{figure*}

\begin{figure*}
\centerline{\includegraphics[width=16.cm]{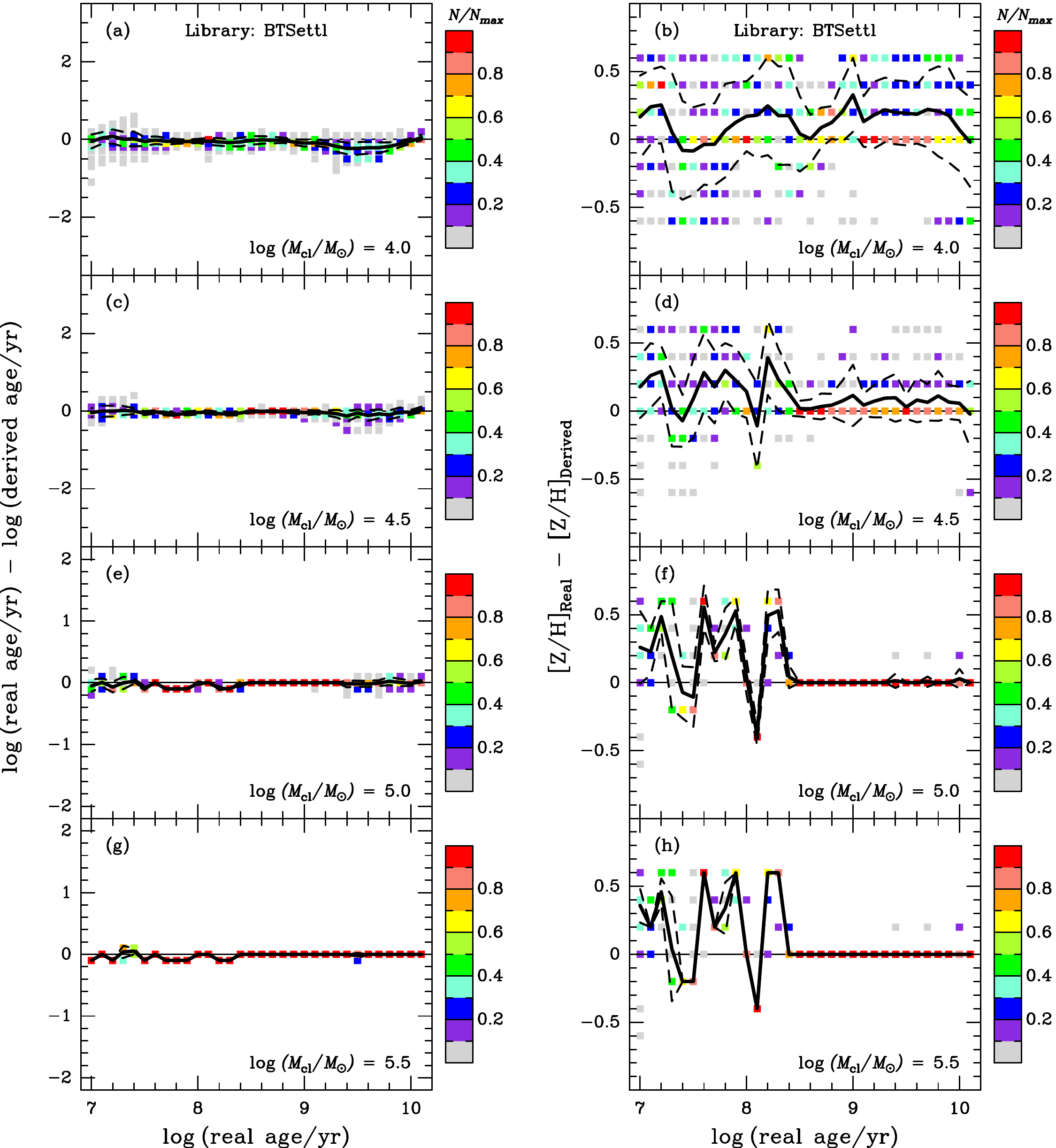}}
\caption{Similar to Figure~\ref{f:offsets_age_ZH_opt}, but now using the BT-Settl library.}
\label{f:offsets_age_ZH_opt_BT}
\end{figure*}

\begin{figure*}
\centerline{\includegraphics[width=16.cm]{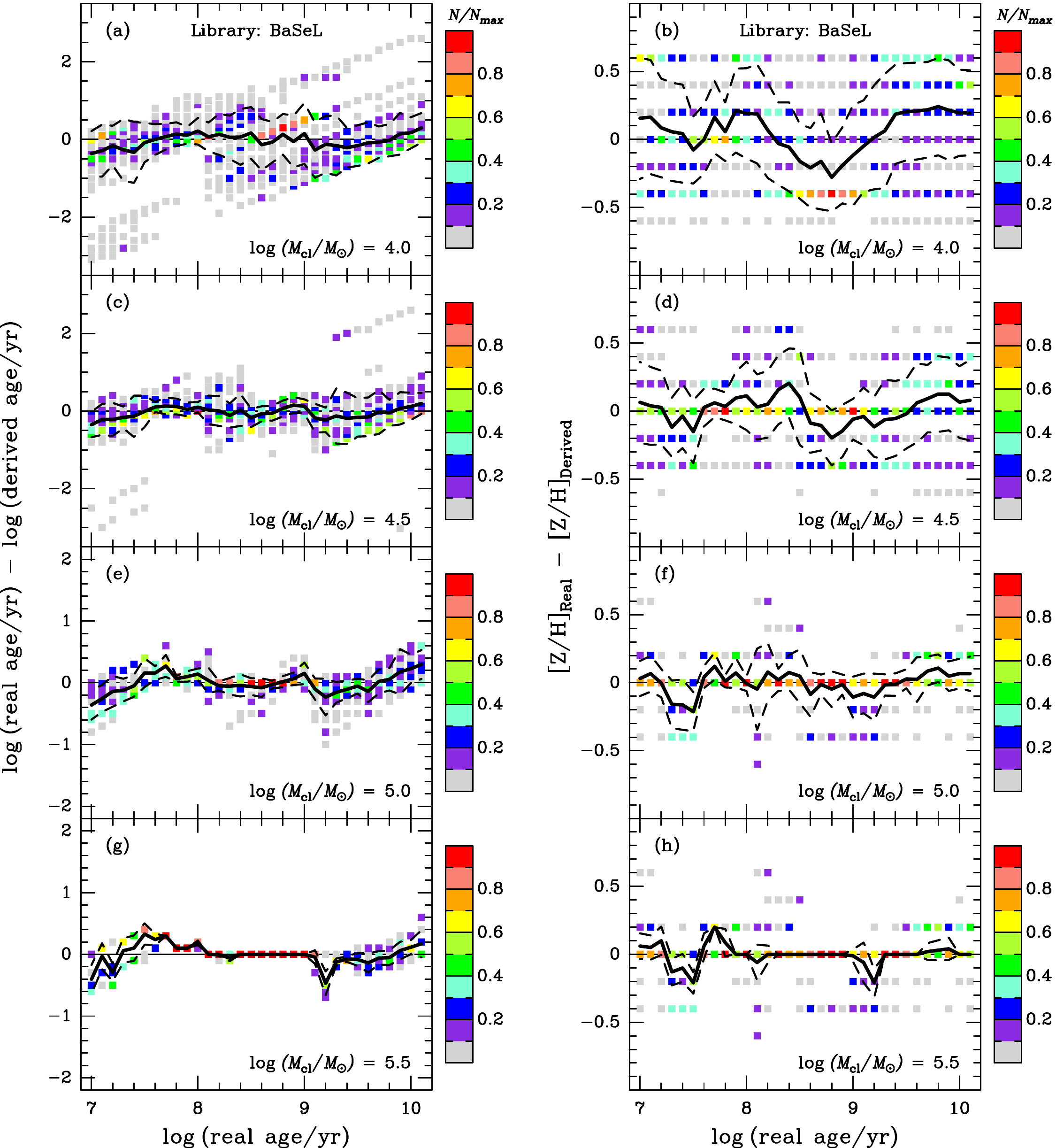}}
\caption{Similar to Figure~\ref{f:offsets_age_ZH_opt}, but now for the red optical range.}
\label{f:offsets_age_ZH_red}
\end{figure*}

\begin{figure*}
\centerline{\includegraphics[width=16.cm]{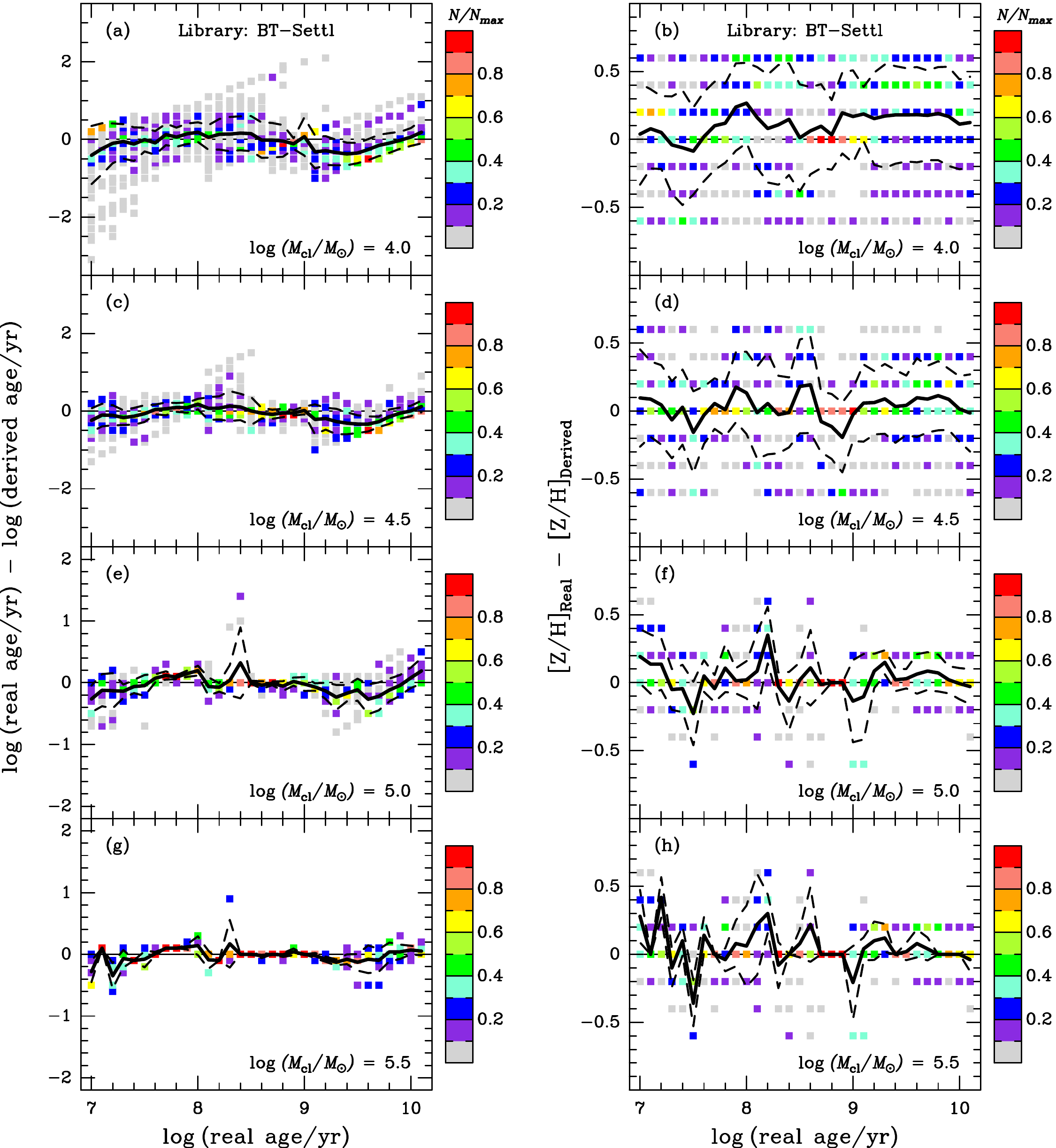}}
\caption{Similar to Figure~\ref{f:offsets_age_ZH_opt_BT}, but now for the red optical range.}
\label{f:offsets_age_ZH_red_BT}
\end{figure*}

\begin{figure*}
\centerline{\includegraphics[width=16.cm]{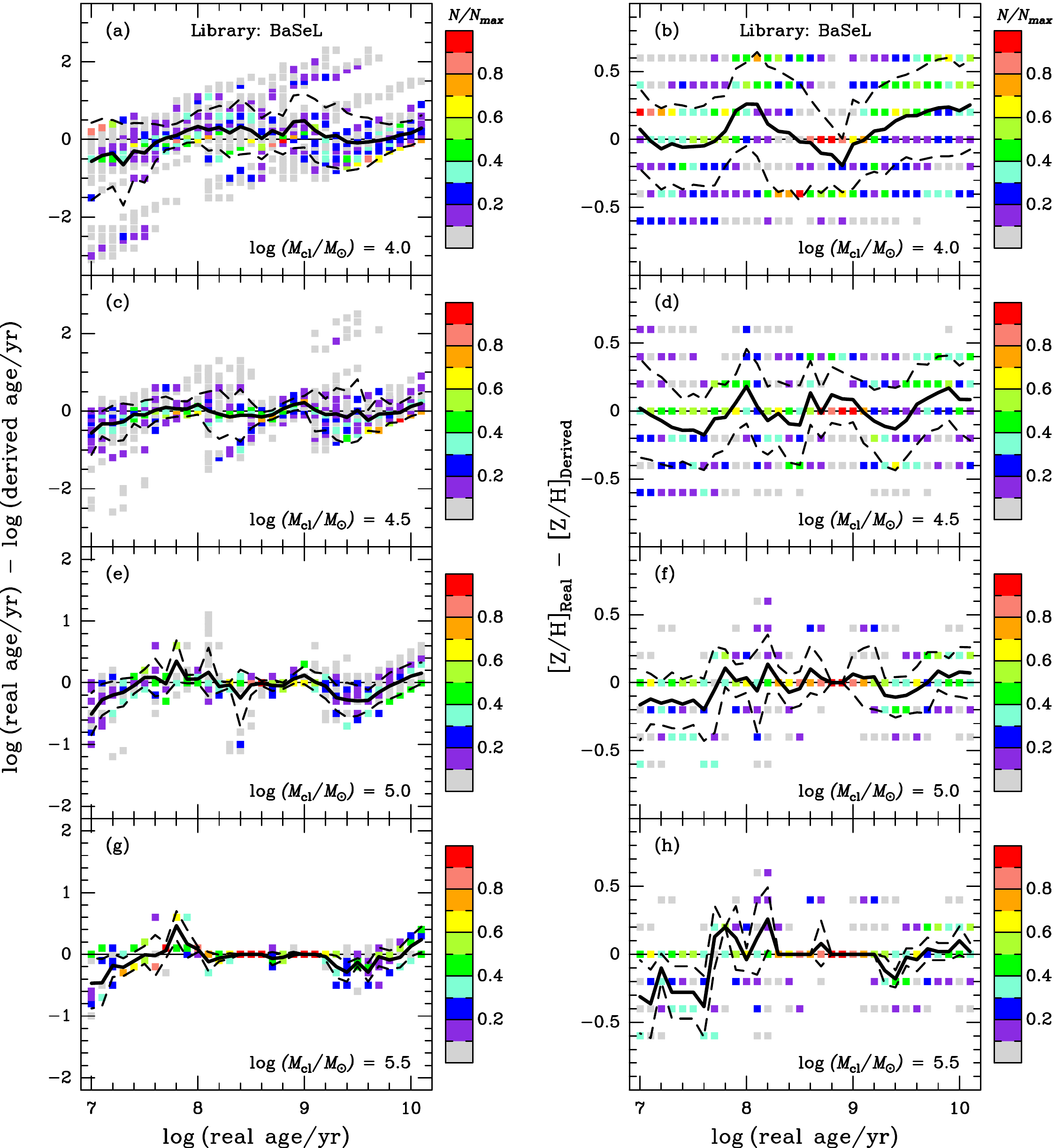}}
\caption{Similar to Figure~\ref{f:offsets_age_ZH_opt}, but now for the NIR range.}
\label{f:offsets_age_ZH_NIR}
\end{figure*}

\begin{figure*}
\centerline{\includegraphics[width=16.cm]{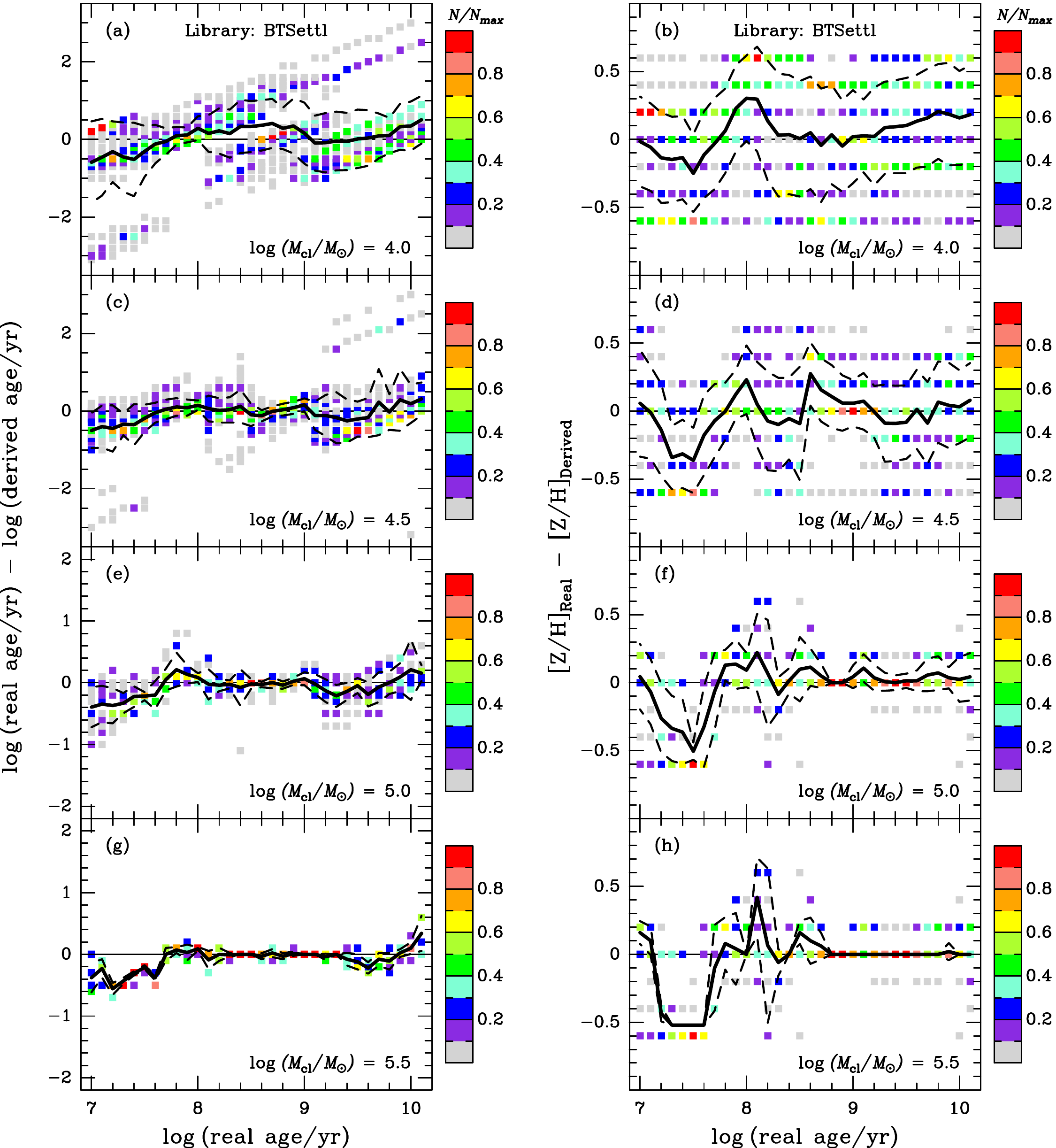}}
\caption{Similar to Figure~\ref{f:offsets_age_ZH_opt_BT}, but now for the NIR range.}
\label{f:offsets_age_ZH_NIR_BT}
\end{figure*}

\begin{figure*}
\centerline{\includegraphics[width=16.cm]{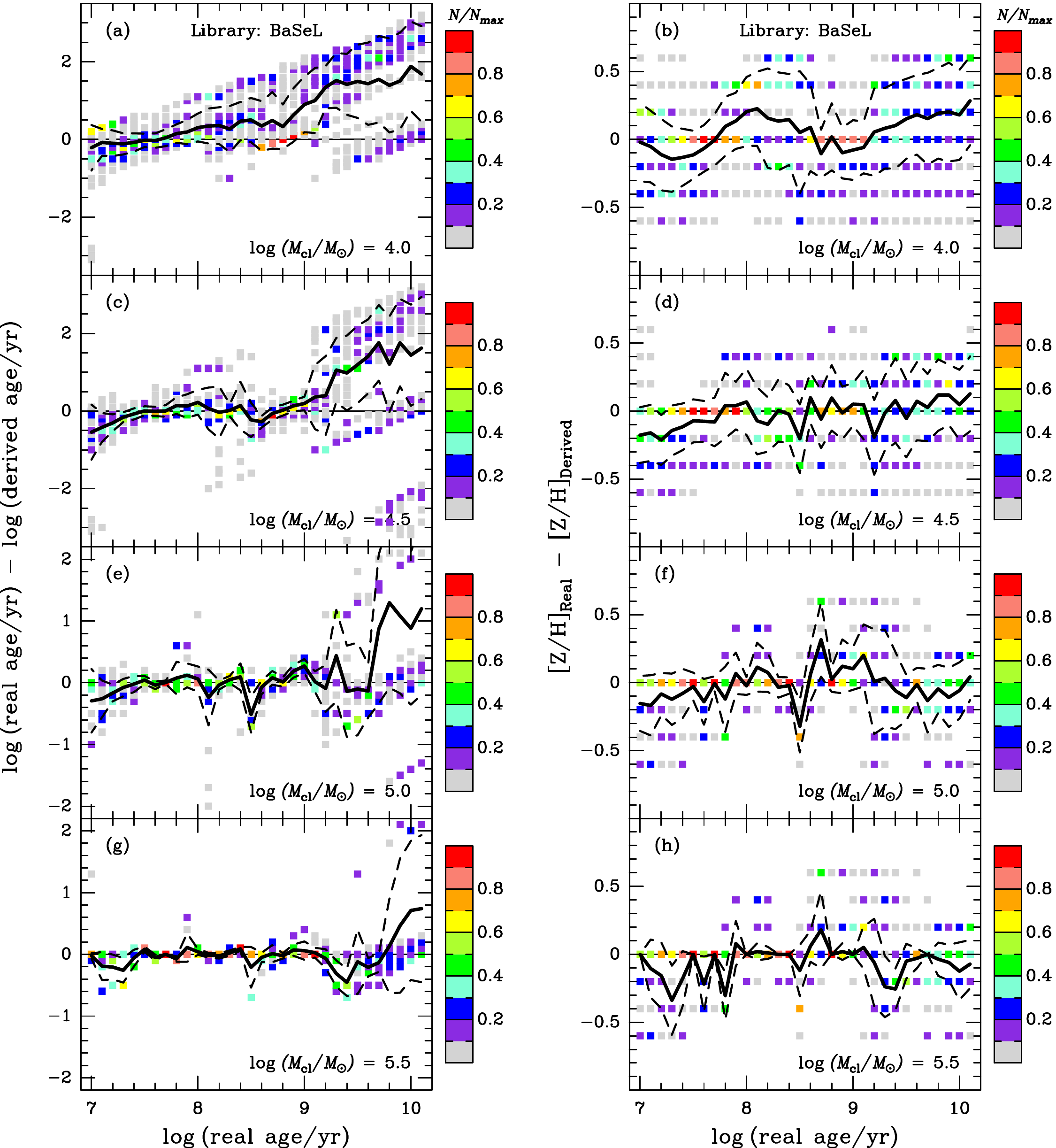}}
\caption{Similar to Figure~\ref{f:offsets_age_ZH_opt}, but now for the mid-IR range.}
\label{f:offsets_age_ZH_midIR}
\end{figure*}

\begin{figure*}
\centerline{\includegraphics[width=16.cm]{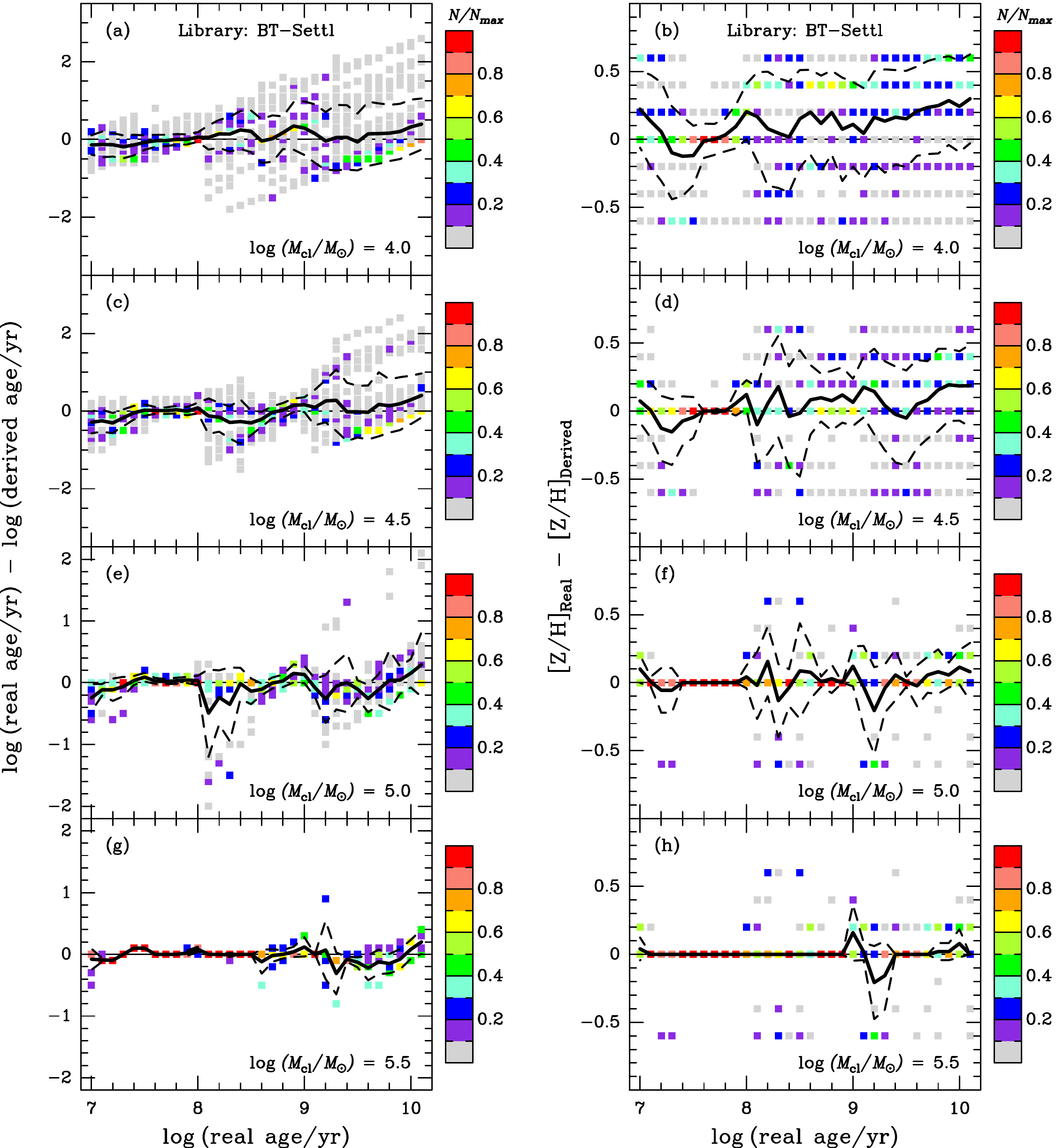}}
\caption{Similar to Figure~\ref{f:offsets_age_ZH_opt_BT}, but now for the mid-IR range.}
\label{f:offsets_age_ZH_midIR_BT}
\end{figure*}

\label{lastpage}

\end{document}